\documentclass[conference]{IEEEtran}

\usepackage{widetext} % Use this to provide widetext environment when using IEEEtran
\usepackage{cite} % Use this to make nicer citations when using IEEEtran
\usepackage{graphicx} % support the \includegraphics command and options
\usepackage{tikz}  
\usetikzlibrary{backgrounds,fit,decorations.pathreplacing,calc,decorations.pathmorphing} 

\usepackage{array} % for better arrays (eg matrices) in maths
\usepackage{verbatim} % adds environment for commenting out blocks of text & for better verbatim
\usepackage{amsmath,amsfonts,amssymb,amscd}
\usepackage{amsthm}
\usepackage{thmtools,thm-restate}

\usepackage{hyperref}
\hypersetup{
    bookmarksnumbered=true, % If Acrobat bookmarks are requested, include section numbers
    unicode=false, % non-Latin characters in Acrobat bookmarks
    pdfstartview={FitH}, % fits the width of the page to the window
    pdftitle={}, % title
    pdfauthor={}, % author
    pdfsubject={}, % subject of the document
    pdfcreator={}, % creator of the document
    pdfproducer={}, % producer of the document
    pdfkeywords={}, % list of keywords
    pdfnewwindow=true, % links in new window
    colorlinks=true, % false: boxed links; true: colored links
    linkcolor=blue, % color of internal links
    citecolor=blue, % color of links to bibliography
    filecolor=blue, % color of file links
    urlcolor=blue % color of external links
}

\newtheorem{thm}{Theorem}[section]

\newtheorem{prop}[thm]{Proposition}

\numberwithin{equation}{section}

\newcommand{\eq}[1]{\hyperref[eq:#1]{Eq. (\ref*{eq:#1})}}

\renewcommand{\sec}[1]{\hyperref[sec:#1]{Section~\ref*{sec:#1}}}
\newcommand{\fig}[1]{\hyperref[fig:#1]{Figure~\ref*{fig:#1}}}

\newcommand{\ket}[1]{|#1\rangle}
\newcommand{\bra}[1]{\langle#1|}

\usepackage{xcolor}

\usepackage{marginnote}

\DeclareMathAlphabet{\matheu}{U}{eus}{m}{n}

\DeclareMathOperator{\tr}{tr}
\DeclareMathOperator{\Tr}{tr}

\newcommand{\sop}[1]{{\mathcal #1}}

\newcommand{\ketbra}[2]{|{#1}\rangle\!\langle{#2}|}

\newcommand{\tensr}{\otimes}

\newcommand{\set}[1]{\lbrace{#1}\rbrace}
\newcommand{\abs}[1]{\lvert{#1}\rvert}
\newcommand{\norm}[1]{\lVert{#1}\rVert}

\newcommand{\EE}{\mathbb{E}}
\newcommand{\CC}{\mathbb C}
\newcommand{\RR}{\mathbb R}

\newcommand{\eps}{\varepsilon}
\DeclareMathOperator{\poly}{poly}
\newcommand{\calA}{\mathcal{A}}
\newcommand{\Herm}{\text{Herm}}
\newcommand{\true}{\text{true}}
\newcommand{\opt}{\text{opt}}

\newcommand{\desU}{C}
\newcommand{\caldes}{\mathcal{C}}

\newcommand{\calC}{\mathcal{C}}

\newcommand{\calE}{\mathcal{E}}

\newcommand{\calI}{\mathcal{I}}
\newcommand{\calL}{\mathcal{L}}
\newcommand{\calU}{\mathcal{U}}
\newcommand{\calV}{\mathcal{V}}
\newcommand{\calY}{\mathcal{Y}}

\newcommand{\tilf}{\tilde f}
\newcommand{\hatf}{\hat f}
\newcommand{\tilA}{\tilde{\sop A}}
\newcommand{\hatA}{\hat{\sop A}}
\newcommand{\tilE}{\tilde{E}}
\newcommand{\hatE}{\hat{ E}}
\newcommand{\tilG}{\tilde{G}}
\newcommand{\hatG}{\hat{ G}}
\newcommand{\tilQ}{\tilde{Q}}
\newcommand{\hatQ}{\hat{ Q}}
\newcommand{\tilW}{\tilde{W}}
\newcommand{\hatW}{\hat{ W}}
\newcommand{\tilH}{\tilde{H}}
\newcommand{\hatH}{\hat{ H}}
\newcommand{\tilS}{\tilde{S}}
\newcommand{\hatS}{\hat{ S}}

\newcommand{\VEC}{\text{vec}}

\begin{document}
\title{Phase Retrieval Using Unitary 2-Designs}
\author{
\IEEEauthorblockN{Shelby Kimmel\IEEEauthorrefmark{1} and 
Yi-Kai Liu\IEEEauthorrefmark{1}\IEEEauthorrefmark{2}}
\IEEEauthorblockA{\IEEEauthorrefmark{1}Joint Center for Quantum Information and Computer Science (QuICS)\\
University of Maryland, College Park, MD, USA}
\IEEEauthorblockA{\IEEEauthorrefmark{2}National Institute of Standards and Technology (NIST)\\
Gaithersburg, MD, USA}
}

\maketitle % Do maketitle before abstract, when using IEEEtran

\begin{abstract}
We consider a variant of the phase retrieval problem, where vectors are replaced by unitary matrices, i.e., the unknown signal is a unitary matrix $U$, and the measurements consist of squared inner products $\abs{\Tr(C^\dagger U)}^2$ with unitary matrices $C$ that are chosen by the observer. This problem has applications to quantum process tomography, when the unknown process is a unitary operation. 

We show that PhaseLift, a convex programming algorithm for phase retrieval, can be adapted to this matrix setting, using measurements that are sampled from \textit{unitary 4- and 2-designs}. In the case of unitary 4-design measurements, we show that PhaseLift can reconstruct all unitary matrices, using a near-optimal number of measurements. This extends previous work on PhaseLift using spherical 4-designs. 

In the case of unitary 2-design measurements, we show that PhaseLift still works pretty well on average: it recovers almost all signals, up to a constant additive error, using a near-optimal number of measurements. These 2-design measurements are convenient for quantum process tomography, as they can be implemented via randomized benchmarking techniques. This is the first positive result on PhaseLift using 2-designs. \end{abstract}

% \maketitle

%%%%%%%%%%%%%%%%%%%%%%%%%%%%%%%%%%%%%%%%%%%%%%%%%%%%%%%%%%%%%%

\section{Introduction}

\subsection{Phase Retrieval for Unitary Matrices}

Phase retrieval is the problem of reconstructing an unknown vector $x \in \CC^d$ from measurements of the form $\abs{\langle a_i, x \rangle}^2$ (for $i = 1,\ldots,m$), where the $a_i \in \CC^d$ are known vectors. This problem has been studied extensively in a number of contexts, including X-ray crystallography and optical imaging \cite{millane90, shechtman15}. 

In this paper we introduce a variant of the phase retrieval problem, where the vectors are replaced by \textit{unitary matrices}: we want to reconstruct an unknown unitary matrix $U \in \CC^{d\times d}$ from measurements of the form $\abs{\Tr(C_i^\dagger U)}^2$ (for $i = 1,\ldots,m$), where the $C_i \in \CC^{d\times d}$ are known unitary matrices. (Here, $C_i^\dagger$ denotes the adjoint of the matrix $C_i$, and $C_i^*$ denotes the complex conjugate.) This problem has applications in quantum process tomography, in the scenario where the experimenter wants to characterize an unknown unitary operation \cite{gross10, liu11, flammia12}.

From a mathematical point of view, this problem can be seen to be a special case of phase retrieval, where the measurements have some additional algebraic structure: in addition to being vectors in $\CC^{d^2}$, they are elements of the $d\times d$ unitary group $U(\CC^{d\times d})$. This can be compared with previous work on phase retrieval, where the measurements were vectors in the unit sphere $S^{d-1} \subset \CC^d$. In particular, a recent line of work has led to tractable algorithms for phase retrieval, for measurements that are \textit{random vectors} in $S^{d-1}$, sampled from the Haar distribution, or from spherical 4-designs \cite{CSV13,GKK15,CL14,KRT15,KZG16}. (We will define these terms more precisely in the following sections.) 

The main contribution of this paper is to prove analogous results when the measurements are \textit{random matrices} in $U(\CC^{d\times d})$, sampled from the Haar distribution, or from unitary 4-designs. In addition, this paper proves weaker recovery guarantees when the measurements are sampled from spherical and unitary 2-designs. (These measurements are of interest because they are easier to implement in experiments, e.g., for quantum process tomography.) In the following sections, we will describe these results in more detail.

As a side note, while we have argued that it is natural to consider measurement matrices $C_i$ that are unitary, one may ask whether it is still possible to recover \textit{all} matrices $U$, and not just unitary ones? Unfortunately, the answer is no. For example, in the $d=2$ case, consider the matrices 
\begin{equation}
U = \begin{pmatrix} 1 & 0 \\ 0 & 0 \end{pmatrix} \text{ and }
V = \begin{pmatrix} 0 & 0 \\ 0 & 1 \end{pmatrix}.
\end{equation}
These matrices cannot be distinguished using any measurement of the form $U \mapsto \abs{\Tr(C^\dagger U)}^2$, where $C \in \CC^{2\times 2}$ is unitary. (To see this, note that any such $C$ can be written in the form 
\begin{equation}
C = \begin{pmatrix} \alpha & e^{i\varphi} \beta^* \\ \beta & -e^{i\varphi} \alpha^* \end{pmatrix},
\end{equation}
where $\alpha, \beta \in \CC$, $\abs{\alpha}^2 + \abs{\beta}^2 = 1$, and $\varphi \in \RR$. Hence we have $\abs{\Tr(C^\dagger U)}^2 = \abs{\alpha}^2 = \abs{\Tr(C^\dagger V)}^2$, i.e., the measurement results are the same for $U$ and $V$.) Thus, the best we can hope for is to reconstruct some large \textit{subset} of matrices in $\CC^{d\times d}$, such as the set of all unitary matrices.

%%%%%%%%%%%%%%%%%%%%%%%%%%%%%%%%%%%%%%%%%%%%%%

\subsection{PhaseLift Algorithm}

PhaseLift is a tractable algorithm for solving the phase retrieval problem, which works by ``lifting'' the quadratic problem in $x$ to a linear problem in $xx^\dagger$ \cite{balan09,CESV13}. We propose the following variant of PhaseLift, for phase retrieval of unitary matrices. 

Suppose we want to recover an unknown unitary matrix $U \in \CC^{d\times d}$. Our approach will be to solve for a Hermitian matrix $\Gamma \in \CC^{d^2\times d^2}_\text{Herm}$, in such a way that the solution has the form 
\begin{equation}
\Gamma_\text{ideal} = \VEC(U)\VEC(U)^\dagger,
\end{equation}
from which we can reconstruct $U$ (up to a global phase factor). Here, $\VEC$ denotes the map $\VEC:\: \CC^{d\times d} \rightarrow \CC^{d^2}$ that ``flattens'' a $d\times d$ matrix into a $d^2$-dimensional vector, 
\begin{equation}
\VEC(U) = (U_{1,1}, U_{1,2}, U_{1,3}, \ldots, U_{d,d-1}, U_{d,d})^T.
\end{equation}
Note that this map preserves inner products: $\langle \VEC(U), \VEC(V) \rangle = \Tr(U^\dagger V)$. 

We can describe our quadratic measurements of $U$ as follows. Let $C_1,\ldots,C_m \in \CC^{d\times d}$ be the unitary measurement matrices introduced earlier. We use the same normalization convention as in previous work \cite{KRT15}: we work with re-scaled matrices $\sqrt{d} C_i$, which have roughly the same magnitude, in Frobenius or $\ell_2$ norm, as Gaussian random matrices.\footnote{To justify this choice, we recall the definition of the Frobenius norm, $\norm{M}_F = \Tr(M^\dagger M)^{1/2} = (\sum_{ij} \abs{M_{ij}}^2)^{1/2}$. For our re-scaled matrices $\sqrt{d} C_i$, we have $\norm{\sqrt{d} C_i}_F^2 = \Tr(\sqrt{d} C_i^\dagger C_i \sqrt{d}) = d^2$. For comparison, we note that a Gaussian random matrix $G \in \CC^{d\times d}$, whose entries $G_{ij} \in \CC$ are sampled independently from a complex Gaussian distribution with mean 0 and variance 1, has expected squared norm $\EE[ \norm{G}_F^2] = d^2$.}

We define the measurement operator $\calA:\: \CC^{d^2\times d^2}_\text{Herm} \rightarrow \RR^m$ as follows:
\begin{equation}
\label{eqn-silver-A}
\calA(\Gamma) = \Bigl[ \VEC(\sqrt{d} C_i)^\dagger \Gamma \VEC(\sqrt{d} C_i) \Bigr]_{i=1}^m.
\end{equation}
Given the unknown matrix $U$, our measurement process returns a vector 
\begin{equation}
\label{eqn-silver-meas}
\begin{split}
y &= \calA\bigl( \VEC(U)\VEC(U)^\dagger \bigr) + \eps \\
 &= \bigl[ d \abs{\Tr(C_i^\dagger U)}^2 \bigr]_{i=1}^m + \eps, 
\end{split}
\end{equation}
where $\eps \in \RR^m$ is an additive noise term. 

Given the measurement results $y$, and an upper-bound $\eta \geq \norm{\eps}_2$ on the strength of the noise (measured using the $\ell_2$ norm), we will then find $\Gamma$ by solving a convex program:
\begin{equation}
\label{eqn-silver}
\begin{split}
\arg & \min_{\Gamma \in \CC^{d^2\times d^2}_\text{Herm}} \Tr(\Gamma) \text{ such that } \\
& \norm{\calA(\Gamma) - y}_2 \leq \eta, \\
& \Gamma \succeq 0, \\
& \Tr_1(\Gamma) = (I/d) \Tr(\Gamma), \\
& \Tr_2(\Gamma) = (I/d) \Tr(\Gamma).
\end{split}
\end{equation}
Here, $\Tr_1(\Gamma) \in \CC^{d\times d}$ and $\Tr_2(\Gamma) \in \CC^{d\times d}$ denote the partial traces over the first and second indices of $\Gamma$, defined as follows. We view $\Gamma\in \CC^{d^2\times d^2}$ as a matrix whose entries $\Gamma_{(a,i),(b,j)}$ are indexed by $(a,i), (b,j) \in \set{1,\ldots,d}^2$. Then $\Tr_1(\Gamma)$ and $\Tr_2(\Gamma)$ are the matrices whose entries are defined by 
\begin{equation}
\Tr_1(\Gamma)_{i,j} = \sum_{a=1}^d \Gamma_{(a,i),(a,j)}, \quad
\Tr_2(\Gamma)_{a,b} = \sum_{i=1}^d \Gamma_{(a,i),(b,i)}.
\end{equation}
% Here, $\Tr_1(\Gamma) \in \CC^{d\times d}$ denotes the partial trace over the first row- and column-indices of $\Gamma$, 
% \begin{equation}
% \Tr_1(\Gamma) = \Bigl[ \sum_{a=1}^d \Gamma_{(a,i),(a,j)} \Bigr]_{i,j=1}^d, 
% \end{equation}
% and similarly, $\Tr_2(\Gamma) \in \CC^{d\times d}$ denotes the partial trace over the second row- and column-indices of $\Gamma$, 
% \begin{equation}
% \Tr_2(\Gamma) = \Bigl[ \sum_{a=1}^d \Gamma_{(i,a),(j,a)} \Bigr]_{i,j=1}^d.
% \end{equation}

The last three constraints in (\ref{eqn-silver}) have the effect of forcing $\Gamma$ to be a linear combination of terms $\VEC(V)\VEC(V)^\dagger$ where the $V$ are unitary.  (This follows from some standard facts in quantum information theory.  Let $\ket{\Phi^+}$ denote the maximally entangled state $\ket{\Phi^+} = \frac{1}{\sqrt{d}} \sum_{i=1}^d \ket{i}\tensr\ket{i} \in \CC^{d^2}$.  Note that $\Gamma$ is proportional to the Jamiolkowski state $J(\calE) = (\calE \tensr \calI)(\ket{\Phi^+}\bra{\Phi^+}) \in \CC^{d^2\times d^2}$ of some quantum process $\calE:\: \CC^{d\times d} \rightarrow \CC^{d\times d}$.  The last two constraints in (\ref{eqn-silver}) imply that $\Tr_1(J(\calE)) = \Tr_2(J(\calE)) = I/d$.  This implies that $\calE$ is unital and trace-preserving, which implies that $\calE$ is an affine combination of unitary processes $\calV$ \cite{Mendl08}.  Hence $J(\calE)$ is an affine combination of terms $J(\calV) = \frac{1}{d} \VEC(V)\VEC(V)^\dagger$, where the $V$ are unitary.)

Also, note that the desired solution $\Gamma_{\text{ideal}} = \VEC(U)\VEC(U)^\dagger$ satisfies these constraints, since we have $\Tr_1(\Gamma_{\text{ideal}}) = (U^\dagger U)^T = I$, $\Tr_2(\Gamma_{\text{ideal}}) = UU^\dagger = I$, and $\Tr(\Gamma_{\text{ideal}}) = \Tr(U^\dagger U) = d$.  

%%%%%%%%%%%%%%%%%%%%%%%%%%%%%%%%%%%%%%%%%%%%

\subsection{Random Measurements and Unitary 4-Designs}

We will consider the situation where the measurement matrices $C_1,\ldots,C_m$ are chosen independently at random from some probability distribution over the unitary group $U(\CC^{d\times d})$. We will be interested in several choices for this distribution over $U(\CC^{d\times d})$. One natural choice is the unitarily-invariant distribution, often called the \textit{Haar distribution}, because this gives a well-defined notion of a ``uniformly random'' unitary matrix. However, for practical purposes, Haar-random measurements are often difficult to implement, when the dimension $d$ is large. 

This motivates us to consider \textit{unitary $t$-designs}, which are distributions that have the same $t$'th-order moments as the Haar distribution. Formally, we say that a distribution $\mathcal{D}$ on $\CC^{d\times d}$ is a unitary $t$-design if 
\begin{equation}
\label{eqn-4th-moment-tensor}
\int_{\mathcal{D}} V^{\tensr t} \tensr (V^\dagger)^{\tensr t} \, dV = 
\int_{\text{Haar}} V^{\tensr t} \tensr (V^\dagger)^{\tensr t} \, dV.
\end{equation}
(Here, $A \tensr B$ denotes the Kronecker product of two matrices $A$ and $B$, and $A^{\tensr t}$ denotes the $t$-fold Kronecker product $A \tensr A \tensr \cdots \tensr A$.) Unitary $t$-designs have been studied in quantum information theory, where they are used to perform tasks such as quantum encryption, fidelity estimation and decoupling \cite{ambainis04, emerson07, ambainis07, scott2008optimizing, dankert09, szehr13}. For many purposes, it is sufficient to use unitary $t$-designs where $t$ is much smaller than $d$, e.g., $t=2$, $t=4$, or $t = \text{poly}(\log d)$. In these cases, there are explicit and computationally efficient constructions for these designs \cite{GAE07,Low10,BHH12,CLLW15,KueGr15,Webb15,Zhu15}.

We will show that the PhaseLift algorithm in equation (\ref{eqn-silver}) succeeds in reconstructing the unknown matrix $U$, when the measurement matrices $C_1,\ldots,C_m$ are chosen at random from a \textit{unitary 4-design}. In particular, we prove a recovery theorem that has a number of attractive features. First, the number of measurements $m$ is close to optimal (up to a factor of $\log d$), since the unknown matrix $U$ has $\Omega(d^2)$ degrees of freedom. Second, the failure probability is exponentially small in $m$, and the resulting recovery guarantee is uniform over all $U$. Third, the recovery is robust to noise, with an explicit bound on the error as a function of $\eta$ and $m$. 

Our proof builds on the work of \cite{KRT15}, who showed an analogous result for PhaseLift when the measurements are random vectors sampled from a spherical 4-design. We use the same high-level approach, known as Mendelson's small ball method \cite{koltchinskii13, mendelson14, T14}. Our main technical innovation is the use of diagrammatic calculus and Weingarten functions \cite{collins03, collins06, scott2008optimizing}, in order to bound the 4th-moment tensor shown in equation (\ref{eqn-4th-moment-tensor}).

Formally, we prove the following bound on the solution that is returned by the PhaseLift algorithm:
\begin{thm}
\label{thm-main-III}
There exists a numerical constant $c_5 > 0$ such that the following holds.
Fix any $c_0 > 2c_5$. 
Consider the above scenario where $m$ measurement matrices
$\desU_1,\ldots,\desU_m$ are chosen independently at random from a unitary
4-design $\hatG$ in $\CC^{d\times d}$.  

Suppose that the number of measurements satisfies 
\begin{equation}
m \geq \bigl( 64(4!)^2 c_0 \bigr)^2 \cdot d^2 \ln d.
\end{equation}
Then with probability at least $1 - \exp\bigl(-2 m \, (4(4!))^{-4}\bigr)$ (over the choice of the $C_i$), we have the following uniform recovery guarantee:  

For any unitary matrix $U \in \CC^{d\times d}$, 
it is the case that any solution $\Gamma_{\opt}$ to the convex program
(\ref{eqn-silver}) with noisy measurements (\ref{eqn-silver-meas}) must
satisfy:
\begin{equation}
\norm{\Gamma_{\opt} - \VEC(U)\VEC(U)^\dagger}_F 
\leq \tfrac{128 (4!)^2 \eta}{\sqrt{m}} \Bigl( 1+\tfrac{2c_5}{c_0-2c_5} \Bigr).
\end{equation}
\end{thm}
We will present the proof in Section \ref{sec:design_4}.

%%%%%%%%%%%%%%%%%%%%%%%%%%%%%%%%%%%%%%%%%%%%

\subsection{Phase Retrieval Using Unitary 2-Designs}
\label{sec-campfire}

Next, we ask the question: how well does PhaseLift perform when the measurement matrices are sampled from a unitary 2-design, rather than a 4-design? 

This question is motivated by a number of practical considerations. First, many experimental methods for quantum tomography make use of random Clifford operations \cite{Gott97, DLT02}, which form a unitary 3-design, but not a 4-design \cite{KueGr15, Webb15, Zhu15}. Moreover, it is generally easier to construct unitary 2-designs, either using quantum circuits or group-theoretic methods \cite{dankert09, CLLW15, GAE07}.  The latter are particularly convenient for randomized benchmarking tomography, where the group structure is essential \cite{K14,JSR15}.

We show that in this situation, PhaseLift still achieves
\textit{approximate} recovery of \textit{almost all} unitary matrices.  
Here, the number of measurements $m$ is still $O(d^2 \text{poly}(\log d))$, which is close to optimal.  
``Approximate recovery'' means that the algorithm recovers the desired solution $\VEC(U)\VEC(U)^\dagger$ 
up to an additive error of size $\delta \norm{\VEC(U)\VEC(U)^\dagger}_F = \delta d$, 
where $\delta \ll 1$ is some constant that is 
independent of the dimension $d$.  We show that this happens for all
unitary matrices $U \in \CC^{d\times d}$, except for a subset that is small with respect to Haar measure.  

In addition, we prove an analogous result for recovery of \textit{vectors} using PhaseLift, when the measurements are sampled from a \textit{spherical} 2-design. Again, we can show approximate recovery of almost all vectors in $\CC^d$, using $m = O(d \text{poly}(\log d))$ measurements. 

These results give a clearer picture of the kinds of measurements that 
are sufficient for PhaseLift to succeed.  In a sense, 2-designs are
\textit{almost} sufficient:  while PhaseLift can sometimes fail using
2-design measurements, our results show that these failures
occur very infrequently.  
(An explicit example of such a failure was previously shown in \cite{GKK15}:  there exists a spherical 2-design $\mathcal{D}$ in $\CC^d$, and there exist vectors $x \in \CC^d$, such that phase retrieval of $x$ requires $m = \Omega(d^2)$ measurements.)

Our proofs require a new ingredient, which is a notion of \textit{non-spikiness} of the unknown vector or matrix that one is trying to recover.  Informally, we say that the unknown vector is ``non-spiky'' if it has small inner product with all of the possible measurement vectors.  (This is reminiscent of previous work on low-rank matrix completion \cite{CR09,gross11,NW12}.)  

Our proofs proceed in two steps:  first, we show that \textit{almost all} vectors are non-spiky, and then we prove that \textit{all} non-spiky vectors can be recovered (uniformly) using PhaseLift.  In the second step, the non-spikiness property plays a crucial role in bounding certain 4th-moment quantities, where we can no longer rely on the 4th moments of the measurement vectors, because the measurements are now being sampled from a 2-design.

We now state our results formally, focusing on the recovery of \textit{matrices}.  (We will describe our results on the recovery of \textit{vectors} in Section \ref{sec-whale}.)  

Let $\tilG$ be a finite set of unitary matrices in $\CC^{d\times d}$.  We say that a unitary matrix $U \in \CC^{d\times d}$ is \textit{non-spiky} with respect to $\tilG$ (with parameter $\beta \geq 0$) if the following holds:
\begin{equation}
\label{eqn-non-spiky-II}
\abs{\Tr(C^\dagger U)}^2 \leq \beta, \quad \forall C \in \tilG.
\end{equation}
Generally speaking, we will say that $U$ is non-spiky when $\beta
\ll d$, e.g., $\beta \leq \poly(\log d)$.
Our first result shows that, when the set $\tilG$ is not too large, almost all unitary matrices $U$ are non-spiky with respect to $\tilG$.
\begin{prop}
\label{prop-non-spiky-II}
Choose $U \in \CC^{d\times d}$ to be a Haar-random unitary matrix.
Then for any $t\geq 0$, with probability at least $1 - 4e^{-t}$ (over
the choice of $U$), $U$ is non-spiky with respect to $\tilG$, with
parameter
\begin{equation}
\beta = \tfrac{9\pi^3}{2} (t + \ln\abs{\tilG}).
\end{equation}
\end{prop}

We will be interested in cases where the vectors in $\tilG$ form a unitary
2-design in $\CC^{d\times d}$.  In these cases, the set $\tilG$ can be
relatively small, i.e., sub-exponential in $d$.  As an example, let $d
= 2^n$, and let $\tilG \subset \CC^{d\times d}$ be the set of Clifford operations on $n$ qubits,
so we have $\abs{\tilG} \leq 2^{2n^2+3n}$ \cite{Gott97, DLT02}.  Then with high probability,
$U$ is non-spiky with respect to $\tilG$, with parameter $\beta =
O(\log^2 d)$.

We then define a non-spiky variant of the PhaseLift algorithm.  This consists of the convex program (\ref{eqn-silver}), together with the following additional constraint, which forces the solution to be non-spiky:
\begin{equation}
\label{eqn-silver-nonspiky}
0 \leq \VEC(C)^\dagger \Gamma \VEC(C) \leq \beta, \quad \forall C \in \tilG.
\end{equation}

We prove that the following recovery guarantee for this algorithm:
\begin{thm}
\label{thm-main-II}
There exists a numerical constant $c_5 > 0$ such that the following holds. 
Fix any $\delta > 0$ and $c_0 > 2c_5$. 
Consider the above scenario where $m$ measurement matrices
$C_1,\ldots,C_m$ are chosen independently at random from a unitary
2-design $\tilG \subset \CC^{d\times d}$.  
Let $\beta \geq 0$, and define $\nu = \beta/\delta$.  

Suppose that the number of measurements satisfies
\begin{equation}
m \geq \bigl( 8c_0 \nu^2 \bigr)^2 \cdot d^2 \ln d.
\end{equation}
Then with probability at least $1-\exp(-\frac{1}{128} m \nu^{-4})$ (over the choice of
the $C_i$), we have the following uniform recovery guarantee:

For any unitary matrix $U \in \CC^{d\times d}$ that is non-spiky with respect to $\tilG$ (with parameter $\beta$, in the sense of 
(\ref{eqn-non-spiky-II})), it is the case that any solution $\Gamma_{\opt}$ 
to the convex program in (\ref{eqn-silver}) and (\ref{eqn-silver-nonspiky}), with noisy measurements (\ref{eqn-silver-meas}), must satisfy:
\begin{multline}
\norm{\Gamma_{\opt} - \VEC(U)\VEC(U)^\dagger}_F \\
\leq \max\Bigl\lbrace \delta \norm{\VEC(U)\VEC(U)^\dagger}_F,\; 
\tfrac{16\eta\nu^2}{\sqrt{m}} \bigl(1 + \tfrac{2c_5}{c_0-2c_5}\bigr) 
\Bigr\rbrace.
\end{multline}
\end{thm}
Note that $m$, the number of measurements, again has almost-linear scaling with $d^2$, which is the number of degrees of freedom in the unknown matrix $U$. In addition, $m$ depends on the dimensionless quantity $\nu = \beta/\delta$, where $\beta$ measures the non-spikiness of the unknown matrix $U$, and $\delta$ controls the accuracy of the solution $\Gamma_{\opt}$. In typical cases, we will have $\beta = O(\text{poly}(\log d))$, and we will choose $\delta$ to be constant, hence we have $\nu = O(\text{poly}(\log d))$.

We will present the proofs of these results for phase retrieval of \textit{matrices} in Section \ref{sec:design_2}, and for phase retrieval of \textit{vectors} in Section \ref{sec-whale}.

%%%%%%%%%%%%%%%%%%%%%%%%%%%%%%%%%%%%%%%%%%%%

\subsection{Application to Quantum Process Tomography}

Finally, we describe an application of our results to quantum process tomography, in the case where the unknown process corresponds to a unitary operation $U \in \CC^{d\times d}$.  This is sufficient to describe the dynamics of a closed quantum system, e.g., time evolution generated by some Hamiltonian, or the action of unitary gates in a quantum computer, including coherent (but not stochastic) errors.  There exist fast methods for learning unitary processes, which achieve a ``quadratic speedup'' over conventional tomography, using ideas from compressed sensing and matrix completion \cite{cramer10, gross10, liu11, shabani11, shabani11b, flammia12, baldwin14, kalev15}.  

Here, we describe an alternative way of achieving this quadratic speedup, using phase retrieval.  Our approach via phase retrieval has a significant advantage:  the measurements can be performed in a way that is robust against state preparation and measurement errors (SPAM errors), by using randomized benchmarking techniques \cite{knill08, magesan11, magesan12, K14, JSR15}.  We discuss this in Section \ref{sec:tomog}.

%%%%%%%%%%%%%%%%%%%%%%%%%%%%%%%%%%%%%%%%%%%%%%%%%%%%%%%%%%%%%%%%

\subsection{Outlook}

In this paper we have studied a variant of the phase retrieval problem that seeks to reconstruct \textit{unitary matrices}, and we have proposed a variant of the PhaseLift algorithm that solves this problem.  We have proved strong reconstruction guarantees when the measurements are sampled from unitary 4-designs, as well as weaker guarantees when the measurements are unitary and spherical 2-designs.  This leads to novel methods for quantum process tomography, when the unknown process is a unitary operation.

We mention a few interesting open problems.  One is to prove error bounds that depend on the $\ell_1$ norm of the noise, rather than the $\ell_2$ norm, as in \cite{CL14}.  Another problem is to extend our recovery method to handle processes with Kraus rank $r>1$ (e.g., stochastic errors), perhaps using the techniques in \cite{KRT15}.

A third problem is to prove tighter bounds when the measurements are random Clifford operations.  Clifford operations are particularly convenient for quantum tomography, and they play an essential role in randomized benchmarking methods.  They are known to be a unitary 3-design, but not a 4-design \cite{KueGr15, Webb15, Zhu15}.  However, our numerical simulations in Section \ref{sec:tomog} seem to indicate that random Clifford measurements perform \textit{better} than their classification as a unitary 3-design would suggest.  This is supported by several recent results showing that random Clifford operations are ``close to'' a unitary 4-design \cite{ZKGG16,HWW16}, and that random vectors chosen from a Clifford orbit perform well for phase retrieval of \textit{vectors} \cite{KZG16}.  Can one prove a similar result for phase retrieval of \textit{matrices}, using random Clifford operations?

Finally, there has been progress in solving phase retrieval problems using gradient descent algorithms, such as Wirtinger Flow \cite{CLS15}.  Can these methods be adapted to our matrix setting?

%%%%%%%%%%%%%%%%%%%%%%%%%%%%%%%%%%%%%%%%%%%%%%%%%%%%%%%%%%%%%%

\section{Outline of the Paper}

In the rest of this paper, we will present the proofs of our theorems, as well as further details on quantum process tomography.  In Section \ref{sec-whale}, we begin with our simplest result, on phase retrieval of \textit{vectors}, using measurements sampled from spherical 2-designs.  In Section \ref{sec:design_2}, we extend this to phase retrieval of unitary \textit{matrices}, using measurements sampled from unitary 2-designs.  In Section \ref{sec:design_4}, we extend this further, to handle measurements sampled from unitary \textit{4-designs}.  Finally, in Section \ref{sec:tomog}, we present further details and numerical simulations regarding quantum process tomography.

\subsection{Notation}

Let $\CC^{d\times d}_{\Herm}$ be the set of $d\times d$ complex
Hermitian matrices.  Let $\calL(\CC^{d\times d}, \CC^{d\times d})$ be
the set of linear maps from $\CC^{d\times d}$ to $\CC^{d\times d}$.
We will use calligraphic letters to denote these maps, e.g., $\calU$,
$\calC$. In general we will consider unitary maps, where
$\calU:\rho\rightarrow U\rho U^\dagger$ for a unitary $U$, and
$\calC:\rho\rightarrow C\rho C^\dagger$ for a unitary $C.$
We will use $\calU$ to represent the unknown map we want to recover,
and $\calC$ to represent the measurement maps we compare with $\calU.$
Thus sometimes $\calC$ will represent an element of a unitary $2$-design,
and sometimes it will represent an element of a unitary $4$-design.

For any matrix $A$, let $A^\dagger$ be its adjoint, let $A^T$ be its
transpose, and let $A^*$ be its complex conjugate.

For any matrix $A$, with singular values $\sigma_1(A) \geq \sigma_2(A)
\geq \cdots \geq \sigma_d(A) \geq 0$, let $\norm{A} = \sigma_1(A)$ be
the operator norm, let $\norm{A}_F = \sqrt{\sum_i \sigma_i^2(A)}$ be
the Frobenius norm, and let $\norm{A}_* = \sum_i \sigma_i(A)$ be the
nuclear or trace norm.

Because we use very similar approaches for the three cases of
spherical 2-designs, unitary 2-designs, and unitary 4-designs,
we will have similar notation in each section. In general,
un-addorned notation (e.g. $f$, $\sop A$) will be used for
spherical 2-designs, notation with tildes (e.g. $\tilf$, $\tilA$)
will be used for unitary 2-designs, and notation with hats
(e.g. $\hatf$, $\hatA$) will be used for unitary 4-designs.

%%%%%%%%%%%%%%%%%%%%%%%%%%%%%%%%%%%%%%%%%%%%%%%%%%%%%%%%%%%%%%

\section{Phase Retrieval and Low-Rank Matrix Recovery Using Spherical 2-Designs}
\label{sec-whale}

We first consider phase retrieval of \textit{vectors}. In fact, we follow the approach of \cite{KRT15}, and consider the more general problem of low-rank matrix recovery. Whereas the authors of \cite{KRT15} showed an exact recovery result for measurements that are sampled from a spherical 4-design, we show an approximate, average-case recovery result for measurements that are chosen from a spherical 2-design.

We want to reconstruct an unknown matrix $X \in
\CC^{d\times d}_{\text{Herm}}$, having rank at most $r$, from
quadratic measurements of the form
\begin{equation}
y_i = w_i^\dagger X w_i + \eps_i, \qquad i=1,2,\ldots,m, 
\end{equation}
where the measurement vectors $w_i \in \CC^d$ are known and are
sampled independently at random from a spherical 2-design, and the
$\eps_i \in \RR$ are unknown contributions due to additive noise.

This problem has been studied previously, particularly in the
rank-1 case (setting $r=1$), where it corresponds to phase retrieval
\cite{GKK15, KRT15}.  Roughly speaking, it is known that
spherical 4-designs are sufficient to recover $X$ efficiently \cite{KRT15}, whereas 
there exists a spherical 2-design that can not recover $X$ efficiently
\cite{GKK15}.  More precisely, $X$ can be recovered (uniformly), via convex
relaxation, from $m = O(rd \poly\log d)$ measurements chosen from any
spherical 4-design; while on the other hand, $X$ cannot be recovered,
by any method, from fewer than $m = \Omega(d^2)$ measurements chosen
from a particular spherical 2-design.

Here, we show that 2-designs are sufficient to recover \textit{a large
subset} of all the rank-$r$ matrices in $\CC^{d\times
d}_{\text{Herm}}$.  More precisely, we show that 2-designs achieve
efficient \textit{approximate} recovery of all low-rank matrices $X$
that are \textit{non-spiky} with respect to the measurement vectors
(we will define this more precisely below).  This implies that
2-designs are sufficient to recover generic (random) low-rank matrices
$X$, since these matrices satisfy the non-spikiness requirement with
probability close to 1.  Here, ``efficient approximate recovery''
means recovery up to an arbitrarily small constant error, using $m =
O(rd \poly\log d)$ measurements, by solving a convex relaxation.  This
is reminiscent of results on non-spiky low-rank matrix completion
\cite{NW12, CR09, gross11}.

%%%%%%%%%%%%%%%%%%%%%%%%%%%%%%%%%%%%%%%%%%%%%%%%%%%%%%%%%%%%%%

\subsection{Non-spikiness condition}

Let $G$ be a finite set of vectors in $\CC^d$, each of length 1.  We say that $X$ is \textit{non-spiky} with respect to $G$ (with parameter $\beta \geq 0$) if the following holds:
\begin{equation}
\label{eqn-non-spiky}
\abs{w^\dagger X w} \leq \frac{\beta}{d} \norm{X}_*, \quad \forall w \in G.
\end{equation}
Here, $\norm{X}_* = \Tr(\abs{X})$ denotes the nuclear norm.  Generally speaking, we will say that $X$ is non-spiky when the parameter $\beta$ is much smaller than $d$, e.g., of size $\poly(\log d)$.  

We now show that, when the set $G$ is not too large, almost all rank-$r$ matrices $X \in \CC^{d\times d}_{\text{Herm}}$ are non-spiky with respect to $G$.  

\begin{prop}
Fix some rank-$r$ matrix $W \in \CC^{d\times d}_{\text{Herm}}$.  Construct a random matrix $X$ by setting $X = UWU^\dagger$, where $U \in \CC^{d\times d}$ is a Haar-random unitary operator.  

Then for any $t\geq 0$, with probability at least $1 - 2e^{-t}$ (over the choice of $U$), $X$ is non-spiky with respect to $G$, with parameter 
\begin{equation}
\beta = 9\pi^3 (t + \ln\abs{G} + \ln r).
\end{equation}
\end{prop}

\noindent
Proof:  Let $S^{d-1}$ denote the unit sphere in $\CC^d$.  Fix any vector $w \in S^{d-1}$, and let $x$ be a uniformly random vector in $S^{d-1}$.  Using Levy's lemma \cite{Matousek02, PSW06}, we have that 
\begin{equation}
\Pr[\abs{w^\dagger x} \geq \eps] \leq 2\exp\bigl( -\tfrac{d\eps^2}{9\pi^3} \bigr).
\end{equation}
Setting $\eps = \sqrt{\tfrac{9\pi^3}{d} (t + \ln\abs{G} + \ln r)}$, we get that 
\begin{equation}
\label{eqn-worm}
\Pr[\abs{w^\dagger x} \geq \eps] \leq 2e^{-t} \abs{G}^{-1} r^{-1}.
\end{equation}

Now recall that $X = UWU^\dagger$, and write the spectral decomposition of $W$ as $W = \sum_{i=1}^r \lambda_i v_i v_i^\dagger$.  Then for any $w \in G$, we can write $w^\dagger X w$ as follows:  
\begin{equation}
\label{eqn-worm2}
w^\dagger X w = w^\dagger U W U^\dagger w = \sum_{i=1}^r \lambda_i \abs{w^\dagger U v_i}^2.
\end{equation}
Each vector $Uv_i$ is uniformly random in $S^{d-1}$, hence it obeys the bound in (\ref{eqn-worm}).  Using the union bound over all $w \in G$ and all $v_1,\ldots,v_r$, we conclude that:  
\begin{equation}
\abs{w^\dagger U v_i} \leq \eps, \quad \forall w \in G,\; \forall i=1,\ldots,r, 
\end{equation}
with failure probability at most $2e^{-t}$.  Combining this with (\ref{eqn-worm2}) proves the claim.  $\square$

\vskip 11pt

We will be interested in cases where the vectors in $G$ form a spherical 2-design in $\CC^d$.  In these cases, the set $G$ can be relatively small, i.e., sub-exponential in $d$.  As an example, let $d = 2^n$, and let $G$ be the set of stabilizer states on $n$ qubits, so we have $\abs{G} \leq 4^{n^2}$ \cite{Gott97}.  Then with high probability, $X$ is non-spiky with respect to $G$, with parameter $\beta = O(\log^2 d)$.

%%%%%%%%%%%%%%%%%%%%%%%%%%%%%%%%%%%%%%%%%%%%%%%%%%%%%%%%%%%%%%

\subsection{Convex relaxation (PhaseLift)}

We now consider measurement vectors $w_1,\ldots,w_m$ that are sampled independently from a spherical 2-design $G \subset \CC^d$.  Using these measurements, we will seek to reconstruct those matrices $X \in \CC^{d\times d}_{\text{Herm}}$ that are low-rank, and non-spiky with respect to $G$.  

We assume we are given a bound on the nuclear norm of $X$, say (without loss of generality): 
\begin{equation}
\norm{X}_* \leq 1.
\end{equation}
As was done in \cite{KRT15}, we will rescale the vectors $w_i$ to be more like Gaussian random vectors, i.e., we will work with the vectors 
\begin{equation}
\label{eqn-mollusc}
a_i = [(d+1)d]^{1/4} w_i, \quad i=1,\ldots,m.
\end{equation}
We also let $A$ be the renormalized version of the spherical 2-design $G$, 
\begin{equation}
A = \set{[(d+1)d]^{1/4} w \;|\; w\in G}.
\end{equation}
Then $X$ will satisfy the following the non-spikiness conditions:
\begin{equation}
\label{eqn-non-spiky2}
\abs{a^\dagger X a} \leq \beta \sqrt{1+\tfrac{1}{d}}, \quad \forall a \in A.
\end{equation}

We define the sampling operator $\calA:\: \CC^{d\times d}_{\text{Herm}} \rightarrow \RR^m$, 
\begin{equation}
\calA(Z) = \bigl( a_i^\dagger Z a_i \bigr)_{i=1}^m.
\end{equation}
Using this notation, the measurement process returns a vector 
\begin{equation}
\label{eqn-meas2}
y = \calA(X) + \eps, \quad \norm{\eps}_2 \leq \eta, 
\end{equation}
where we assume we know an upper-bound $\eta$ on the size of the noise term.  

We will solve the following convex relaxation, which is essentially the PhaseLift convex program, augmented with non-spikiness constraints:
\begin{equation}
\label{eqn-convex}
\begin{split}
\arg &\min_{Z \in \CC^{d\times d}_{\text{Herm}}} \norm{Z}_* \text{ such that} \\
&\norm{\calA(Z)-y}_2 \leq \eta, \\
&\abs{a^\dagger Z a} \leq \beta \sqrt{1+\tfrac{1}{d}}, \quad \forall a \in A.
\end{split}
\end{equation}

%%%%%%%%%%%%%%%%%%%%%%%%%%%%%%%%%%%%%%%%%%%%%%%%%%%%%%%%%%%%%%

\subsection{Approximate recovery of non-spiky low-rank matrices}

We show the following uniform recovery guarantee for the convex relaxation shown in equation (\ref{eqn-convex}):
\begin{thm}
\label{thm-main}
Fix any $\delta > 0$ and $C_0 > 13$.  Consider the above scenario where $m$ measurement vectors $a_1,\ldots,a_m$ are chosen independently at random from a (renormalized) spherical 2-design $A \subset \CC^d$.  Let $1\leq r\leq d$ and $\beta \geq 0$, and define $\nu = 2\beta/\delta$.  

Suppose that the number of measurements satisfies 
\begin{equation}
m \geq \bigl( 8C_0 \nu^2 (1+\tfrac{1}{d}) \bigr)^2 \cdot rd\log(2d).
\end{equation}
Then with probability at least 
\begin{equation*}
1 - \exp\Bigl( -\tfrac{1}{128} \, m \, \bigl( \nu^2 (1+\tfrac{1}{d}) \bigr)^{-2} \Bigr)
\end{equation*}
(over the choice of the $a_i$), we have the following uniform recovery guarantee:  

For any rank-$r$ matrix $X_{\true} \in \CC^{d\times d}_{\text{Herm}}$ that has nuclear norm $\norm{X_{\true}}_* \leq 1$, and is non-spiky with respect to $A$ (with parameter $\beta$, in the sense of (\ref{eqn-non-spiky2})), it is the case that any solution $X_{\opt}$ to the convex program (\ref{eqn-convex}) with noisy measurements (\ref{eqn-meas2}) must satisfy:
\begin{equation}
\begin{split}
&\norm{X_{\opt}-X_{\true}}_F \\
&\quad\leq \max\Bigl\lbrace \delta,\; \tfrac{16\eta \nu^2}{\sqrt{m}} (1+\tfrac{1}{d}) (1+\tfrac{13}{C_0-13}) \Bigr\rbrace.
\end{split}
\end{equation}
\end{thm}

This error bound has similar scaling to the one in \cite{KRT15}:  the number of measurements $m$ is only slightly larger than the number of degrees of freedom $O(rd)$, and the error $X_{\opt} - X_{\true}$ decreases like $\eta/\sqrt{m}$, until it reaches the limit $\delta$.  However, now both of these bounds also involve the dimensionless quantity $\nu$, which in turn depends on the non-spikiness $\beta$ of the original matrix $X_{\true}$, as well as the desired accuracy $\delta$ of the reconstructed matrix $X_{\opt}$.  In some sense, this is the price that one pays when one uses a weaker measurement ensemble, such as a spherical 2-design.  It is an interesting question whether one can improve these bounds, to have a better dependence on $\nu$.

% \textbf{Remark (FIXME):}  \textit{I think this result doesn't have the optimal scaling of $m$ as a function of $\delta$.  From looking at covering numbers of the nuclear-norm ball with respect to Frobenius norm, I'd expect something like $m \sim 1/\delta^2$, but this is really just a wild guess.  However, I don't think this issue prevents the result from being interesting.  I'm more concerned about whether there could be other issues or things that are wrong with this result?  Could this result end up being trivial or obvious for some reason?}

The proof follows the same strategy used in \cite{KRT15} to show low-rank matrix recovery using spherical 4-design measurements, by means of Mendelson's small-ball method \cite{T14}.  The key difference is that our measurements are spherical 2-designs only, so we do not have control over their fourth moments.  Instead, we use the non-spikiness properties of $X_{\true}$ and $X_{\opt}$ to bound the fourth moments that appear in the proof.  This allows us to show approximate recovery of $X_{\true}$, up to an arbitrarily small constant error $\delta$.

We will present this proof in several steps.

%%%%%%%%%%%%%%%%%%%%%%%%%%%%%%%%%%%%%%%%%%%%%%%%%%%%%%%%%%%%%%

\subsection{Approximate recovery via modified descent cone}

We begin by defining a modified version of the descent cone used in
\cite{KRT15,T14}.  Let $f:\: \RR^d \rightarrow \RR \cup \set{\infty}$
be a proper convex function, and let $x_{\true} \in \RR^d$ and $\delta
\geq 0$.  Then we define the modified descent cone $D'(f, x_{\true},
\delta)$ as follows:
\begin{equation}
\begin{split}
D'(f, x_{\true}, \delta) = \lbrace y \in \RR^d \;|\; &\exists \tau>0 \text{ such that } \\
&f(x_{\true}+\tau y) \leq f(x_{\true}),\; \\
&\norm{\tau y}_2 \geq \delta \rbrace.
\end{split}
\end{equation}
This is the set of all directions, originating at the point
$x_{\true}$, that cause the value of $f$ to decrease, when one takes a
step of size at least $\delta$.  Note that this is a cone, but not
necessarily a convex cone.

We use this to state an approximate recovery bound, analogous to Prop.
7 in \cite{KRT15} and Prop. 2.6 in \cite{T14}.  Let $y$ be a noisy
linear measurement of $x_{\true}$, given by $y = \Phi x_{\true} +
\eps$, where $\Phi \in \RR^{m\times d}$ and $\norm{\eps}_2 \leq \eta$.
Then let $x_{\opt}$ be a solution of the convex program
\begin{equation}
\arg \min_{z \in \RR^d} f(z) \text{ such that } \norm{\Phi z - y}_2 \leq \eta.
\end{equation}
Then we have the following recovery bound: 
\begin{equation}
\norm{x_{\opt}-x_{\true}}_2 \leq \max\biggl\lbrace \delta,\; \frac{2\eta}{\lambda_{\min}(\Phi; D'(f, x_{\true}, \delta))} \biggr\rbrace, 
\end{equation}
where $\lambda_{\min}$ is the conic minimum singular value, 
\begin{multline}
\lambda_{\min}(\Phi; D'(f, x_{\true}, \delta)) \\
= \inf\set{\norm{\Phi u}_2 \;|\; u \in S^{d-1} \cap D'(f, x_{\true}, \delta)}.
\end{multline}
This is proved easily, using the same argument as in \cite{KRT15, T14}.

We now re-state this bound for the setup in Theorem \ref{thm-main}.
Here the function $f:\: \CC^{d\times d}_{\text{Herm}} \rightarrow \RR
\cup \set{\infty}$ is given by
\begin{equation}
f(Z) = \begin{cases}
\norm{Z}_* &\text{ if $z$ is non-spiky in} \\
&\textrm{ the sense of (\ref{eqn-non-spiky2}),} \\
\infty &\text{ otherwise,}
\end{cases}
\end{equation}
and we use
\begin{equation}
\begin{split}
D(f, X_{\true}, \delta) = \lbrace Y \in &\CC^{d\times d}_{\text{Herm}} 
\;|\; \exists \tau>0 \text{ such that } \\
&f(X_{\true}+\tau Y) \leq f(X_{\true}),\; \\
&\norm{\tau Y}_F \geq \delta \rbrace.
\end{split}
\end{equation}

Our recovery bound is: 
\begin{equation}
\label{eqn-narwhal}
\norm{X_{\opt}-X_{\true}}_F \leq \max\biggl\lbrace \delta,\; \frac{2\eta}{\lambda_{\min}(\calA; D(f, X_{\true}, \delta))} \biggr\rbrace, 
\end{equation}
and we want to lower-bound the quantity $\lambda_{\min}$: 
\begin{equation}
\label{eqn-dolphin}
\begin{split}
\lambda_{\min}&(\calA; D(f, X_{\true}, \delta)) \nonumber\\
&= \inf_{Y \in E(X_{\true})} \biggl[ \sum_{i=1}^m \abs{\Tr(a_i a_i^\dagger Y)}^2 \biggr]^{1/2}, 
\end{split}
\end{equation}
where we define 
\begin{equation}
\label{eqn-EXtrue}
E(X_{\true}) = \set{Y \in D(f, X_{\true}, \delta) \;|\; \norm{Y}_F = 1}.
\end{equation}

%%%%%%%%%%%%%%%%%%%%%%%%%%%%%%%%%%%%%%%%%%%%%%%%%%%%%%%%%%%%%%

\subsection{Mendelson's small-ball method}

In order to lower-bound $\lambda_{\min}$, we will use Mendelson's
small-ball method, following the steps described in \cite{KRT15} (see
also \cite{T14}).  We will prove a uniform lower-bound that holds for
all $X_{\true}$ simultaneously, i.e., we will lower-bound $\inf_{X_{\true}} \lambda_{\min}$.  We define
\begin{equation}
\label{eqn-E}
E = \bigcup_{X_{\true}} E(X_{\true}), 
\end{equation}
where the union runs over all $X_{\true} \in \CC^{d\times
d}_{\text{Herm}}$ that have rank at most $r$ and satisfy the non-
spikiness conditions (\ref{eqn-non-spiky2}).  We then take the infimum
over all $Y \in E$.

Using Mendelson's small-ball method (Theorem 8 in \cite{KRT15}), we have that
for any $\xi > 0$ and $t \geq 0$, with probability at least $1 - e^{-2t^2}$,
\begin{multline}
\label{eqn-porpoise}
\inf_{Y \in E} \biggl[ \sum_{i=1}^m \abs{\Tr(a_i a_i^\dagger Y)}^2 \biggr]^{1/2} \\
\geq \xi \sqrt{m} Q_{2\xi}(E) - 2W_m(E) - \xi t, 
\end{multline}
where we define 
\begin{align}
Q_\xi(E) &= \inf_{Y\in E} \left\{\Pr_{a\sim \calA}[\abs{\Tr(a a^\dagger Y)} \geq \xi]\right\}, \\
W_m(E) &= \underset{\substack{\epsilon_i\sim\pm1\\a_i\sim \sop A}}{\EE} \left[
\sup_{Y\in E} \Tr(HY)\right], \quad H = \frac{1}{\sqrt{m}} \sum_{i=1}^m \eps_i a_i a_i^\dagger, 
\end{align}
and $\eps_1,\ldots,\eps_m$ are Rademacher random variables.  

%%%%%%%%%%%%%%%%%%%%%%%%%%%%%%%%%%%%%%%%%%%%%%%%%%%%%%%%%%%%%%

\subsection{Lower-bounding $Q_\xi(E)$}

We will lower-bound $Q_\xi(E)$ as follows.  Fix some $Y \in E$.  We
know that $Y$ is in $E(X_{\true})$, for some $X_{\true} \in
\CC^{d\times d}_{\text{Herm}}$ that has rank at most $r$ and is 
non-spiky in the sense of (\ref{eqn-non-spiky2}).  Hence we know that
$\norm{Y}_F = 1$, and there exists some $\tau > 0$ such that
$X_{\true} + \tau Y$ is non-spiky in the sense of 
(\ref{eqn-non-spiky2}), and we have
\begin{equation}
\norm{X_{\true} + \tau Y}_* \leq \norm{X_{\true}}_*, \quad 
\norm{\tau Y}_F \geq \delta.  
\end{equation}
Note that we have $\norm{\tau Y}_F = \tau \geq \delta$.  

We will lower-bound $\Pr[\abs{\Tr(a a^\dagger Y)} \geq \xi]$, for any
$\xi \in [0,1]$, using the Paley-Zygmund inequality, and appropriate
bounds on the second and fourth moments of $\Tr(a a^\dagger Y)$.  Let
us define a random variable 
\begin{equation}
S = |\Tr(a a^\dagger \tau Y)|.  
\end{equation}
We can lower-bound $\EE(S^2)$, using the same calculation as in Prop. 12 of
\cite{KRT15}:
\begin{equation}
\EE(S^2) = \Tr(\tau Y)^2 + \norm{\tau Y}_F^2 \geq 0 + \tau^2.
\end{equation}

To handle $\EE(S^4)$, we need to use a different argument from the one in
\cite{KRT15}, since our $a$ is sampled from a spherical 2-design, not
a 4-design.  Our solution is to upper-bound $S$, using the 
non-spikiness properties of $X_{\true}$ and $X_{\true} + \tau Y$:

\begin{equation}
\begin{split}
S &= \bigl\lvert -\Tr(a a^\dagger X_{\true}) + \Tr(a a^\dagger (X_{\true} + \tau Y)) \bigr\rvert \\
&\leq (2\beta) \sqrt{1+\tfrac{1}{d}}.
\end{split}
\end{equation}
This implies an upper-bound on $\EE(S^4)$:  
\begin{equation}
\EE(S^4) \leq (2\beta)^2 (1+\tfrac{1}{d}) \EE(S^2).
\end{equation}

Putting it all together, we have:  
\begin{equation}
\begin{split}
\underset{a\sim \sop A}{\Pr}[|\Tr(a &a^\dagger Y)| \geq \xi] \\
&= \Pr[S^2 \geq \tau^2\xi^2] \\
&\geq \Pr[S^2 \geq \xi^2\EE(S^2)] \\
&\geq (1-\xi^2)^2 \frac{\EE(S^2)^2}{\EE(S^4)} \\
&\geq (1-\xi^2)^2 \frac{\tau^2}{(2\beta)^2 (1+\tfrac{1}{d})}.
\end{split}
\end{equation}
Finally, 
using the fact that $\tau \geq \delta$, we have that 
\begin{equation}
\label{eqn-Q}
Q_{1/2}(E) \geq (1-\xi^2)^2 \frac{\delta^2}{(2\beta)^2 (1+\tfrac{1}{d})}.
\end{equation}

%%%%%%%%%%%%%%%%%%%%%%%%%%%%%%%%%%%%%%%%%%%%%%%%%%%%%%%%%%%%%%

\subsection{Upper-bounding $W_m(E)$}

Next, we will upper-bound $W_m(E)$, using essentially the same
argument as in \cite{KRT15}.  Recall that the argument in \cite{KRT15}
showed an upper-bound on $W_m(F)$, where $F$ was a slightly different
set than our $E$, and the $a_i$ were sampled from a spherical 4-design
rather than a 2-design.  The argument used the following steps:
\begin{equation}
\begin{split}
W_m(F) &= \EE \sup_{Y\in F} \Tr(HY) \\
&\leq \EE\; 2\sqrt{r} \norm{H} \\
&\leq 2\sqrt{r} \cdot (3.1049) \sqrt{d\log(2d)}, 
\end{split}
\end{equation}
using Lemma 10 and Prop. 13 in \cite{KRT15}, and provided that $m \geq 2d\log d$.

The first step still works to upper-bound $W_m(E)$, because $E \subset
F$.  To see this, recall that the set $F$ was defined as follows:
\begin{equation}
F = \bigcup_{X_{\true}} F(X_{\true}), 
\end{equation}
where the union was over all $X_{\true} \in \CC^{d\times d}_{\text{Herm}}$ with rank at most $r$, and 
\begin{equation}
F(X_{\true}) = \set{Y \in D(\norm{\cdot}_*, X_{\true}, 0) \;|\; \norm{Y}_F = 1}.
\end{equation}
This can be compared with our definition of the set $E$ in
(\ref{eqn-E}) and (\ref{eqn-EXtrue}).

The second step still works for our choice of the $a_i$, because Prop.
13 in \cite{KRT15} does not use the full power of the spherical
4-design.  In fact it only requires that the $a_i$ are sampled from a
spherical 1-design (because the proof relies mainly on the Rademacher
random variables $\eps_i$).  Thus we conclude that
\begin{equation}
\label{eqn-W}
W_m(E) \leq 2\sqrt{r} \cdot (3.1049) \sqrt{d\log(2d)}.
\end{equation}

%%%%%%%%%%%%%%%%%%%%%%%%%%%%%%%%%%%%%%%%%%%%%%%%%%%%%%%%%%%%%%

\subsection{Final result}

Combining equations (\ref{eqn-dolphin}), (\ref{eqn-porpoise}),
(\ref{eqn-Q}) and (\ref{eqn-W}), and setting $\xi = \tfrac{1}{4}$, 
$\nu = 2\beta/\delta$ and $t = \tfrac{1}{16} \sqrt{m} 
\bigl( \nu^2 (1+\tfrac{1}{d}) \bigr)^{-1}$, we get that:
\begin{equation}
\begin{split}
\inf_{X_{\true}} &\lambda_{\min}(\calA; D(f, X_{\true}, \delta)) \\
&\geq \tfrac{1}{4} \sqrt{m} \; \frac{9}{16\nu^2 (1+\tfrac{1}{d})} \\
&\quad - (12.44) \sqrt{rd\log(2d)} - \tfrac{1}{4} t \\
&= \tfrac{1}{8} \sqrt{m} \; \frac{1}{\nu^2 (1+\tfrac{1}{d})} \\
&\quad - (12.44) \sqrt{rd\log(2d)}.
\end{split}
\end{equation}
Now we set $m \geq \bigl( 8C_0 \nu^2 (1+\tfrac{1}{d}) \bigr)^2 \cdot rd\log(2d)$, 
which implies 
\begin{equation}
\frac{\sqrt{m}}{8C_0 \nu^2 (1+\tfrac{1}{d})} \geq \sqrt{rd\log(2d)}, 
\end{equation}
hence 
\begin{equation}
\begin{split}
\inf_{X_{\true}} &\lambda_{\min}(\calA; D(f, X_{\true}, \delta)) \\
&\geq \tfrac{1}{8} \sqrt{m} \; \frac{1}{\nu^2 (1+\tfrac{1}{d})} \; \tfrac{C_0-13}{C_0}.
\end{split}
\end{equation}
This can be plugged into our approximate recovery bound (\ref{eqn-narwhal}).  
This finishes the proof of Theorem \ref{thm-main}.  $\square$

%%%%%%%%%%%%%%%%%%%%%%%%%%%%%%%%%%%%%%%%%%%%%%%%%%%%%%%%%%%%%%
%%%%%%%%%%%%%%%%%%%%%%%%%%%%%%%%%%%%%%%%%%%%%%%%%%%%%%%%%%%%%%

\section{Phase Retrieval Using Unitary 2-Designs}\label{sec:design_2}

Next, we extend our results from \textit{vectors} to unitary \textit{matrices}. First, we consider phase retrieval using measurements that are sampled from unitary 2-designs, as described in Section \ref{sec-campfire}.

\subsection{Non-spikiness}

First, let $\tilG \subset \CC^{d\times d}$ be a unitary 2-design. we prove Proposition \ref{prop-non-spiky-II}, showing that almost all unitary matrices are non-spiky with respect to $\tilG$:

\vskip 11pt

\noindent
Proof:  Fix any $C \in \tilG$.  Using Levy's lemma \cite{L09}, and noting
that the function $U \mapsto \Tr(U^\dagger C)$ has Lipshitz
coefficient $\eta \leq \norm{C}_F = \sqrt{d}$, we have that
\begin{equation}
\Pr[\abs{\Tr(U^\dagger C)} \geq \eps] \leq 4\exp\bigl( -\tfrac{2}{9\pi^3} \eps^2 \bigr).
\end{equation}
Setting $\eps = \sqrt{\tfrac{9\pi^3}{2} (t + \ln\abs{\tilG})}$, we get that 
\begin{equation}
\label{eqn-nematode}
\Pr[\abs{\Tr(U^\dagger C)} \geq \eps] \leq 4e^{-t} \abs{\tilG}^{-1}.
\end{equation}
Using the union bound over all $C \in \tilG$, we conclude that:  
\begin{equation}
\abs{\Tr(U^\dagger C)} \leq \eps, \quad \forall C \in \tilG, 
\end{equation}
with failure probability at most $4e^{-t}$.  This proves the claim.  $\square$

%%%%%%%%%%%%%%%%%%%%%%%%%%%%%%%%%%%%%%%%%%%%%%%%%%%%%%%%

\subsection{Approximate recovery of non-spiky unitary matrices}

Next, let $\tilA$ denote the measurement operator defined in equation (\ref{eqn-silver-A}), where the measurement matrices $C_1,\ldots,C_m$ are chosen independently at random from the unitary 2-design $\tilG$.  

We now prove Theorem \ref{thm-main-II}, showing that all non-spiky unitary matrices can be recovered approximately via PhaseLift, given these measurements. We will present this proof in several steps, following the same strategy as in the previous section.

We will use the following notation.  For any unitary matrix $U \in \CC^{d\times d}$, let $\calU:\: \CC^{d\times d} \rightarrow \CC^{d\times d}$ be the quantum process whose action is given by $\calU:\: \rho \mapsto U\rho U^\dagger$.  Let $J(\calU)$ be the corresponding Jamiolkowski state, as defined in equation (\ref{eqn-J-U}), which can be written as 
\begin{equation}
J(\calU) = \frac{1}{d} \VEC(U)\VEC(U)^\dagger \in \CC^{d^2\times d^2}.
\end{equation}

Using this notation, we can write the desired solution of the PhaseLift convex program as $J(\calU)d$, and we can write the measurement operator $\tilA$ as 
\begin{equation}
\tilA(\Gamma) = \Bigl[ \Tr(\Gamma J(\calC_i) d^2) \Bigr]_{i=1}^m.
\end{equation}
We want to recover the solution $J(\calU)d$, up to an additive error of size $\delta \norm{J(\calU)d}_F = \delta d$.

%%%%%%%%%%%%%%%%%%%%%%%%%%%%%%%%%%%%%%%%%%%%%%%%%%%%%%%%

\subsection{Modified descent cone}

We define the function $\tilf:\: \CC^{d^2\times d^2}_{\Herm} \rightarrow \RR \cup \set{\infty}$ as follows:
\begin{equation}
\label{eqn-f-II}
\tilf(\Gamma) = \begin{cases}
\Tr(\Gamma) &\text{ if $\Gamma$ is non-spiky in the sense } \\
              &\text{ of (\ref{eqn-silver-nonspiky}), and also satisfies the } \\
              &\text{ last three constraints in (\ref{eqn-silver}), } \\
\infty &\text{ otherwise.}
\end{cases}
\end{equation}
Our recovery bound is: 
\begin{multline}
\label{eqn-narwhal-II}
\norm{\Gamma_{\opt} - J(\calU)d}_F \\
\leq \max\biggl\lbrace \delta d,\; \frac{2\eta}{\lambda_{\min}(\tilA; D(\tilf, J(\calU)d, \delta d))} \biggr\rbrace, 
\end{multline}
and we want to lower-bound the quantity $\lambda_{\min}$: 
\begin{equation}
\label{eqn-dolphin-II}
\begin{split}
\lambda_{\min}&(\tilA; D(\tilf, J(\calU)d, \delta d)) \\
&= \inf_{Y \in \tilE(\calU)} \biggl[ \sum_{i=1}^m \abs{\Tr(Y J(\calC_i) d^2)}^2 \biggr]^{1/2}, 
\end{split}
\end{equation}
where we define 
\begin{equation}
\label{eqn-EXtrue-II}
\tilE(\calU) = \set{Y \in D(\tilf, J(\calU)d, \delta d) \;|\; \norm{Y}_F = 1}.
\end{equation}

%%%%%%%%%%%%%%%%%%%%%%%%%%%%%%%%%%%%%%%%%%%%%%%%%%%%%%%%

\subsection{Mendelson's small-ball method}

We will use Mendelson's small-ball method to lower-bound
$\lambda_{\min}$.  We will prove a uniform lower-bound that holds for
all $\calU$ simultaneously, i.e., we will lower-bound $\inf_{\calU} \lambda_{\min}$.  We define
\begin{equation}
\label{eqn-E-II}
\tilE = \bigcup_{\calU} \tilE(\calU), 
\end{equation}
where the union runs over all unitary processes $\calU:\: \rho \mapsto U\rho U^\dagger$ 
such that $U \in \CC^{d\times d}$ satisfies the non-spikiness conditions (\ref{eqn-non-spiky-II}).  
We then take the infimum over all $Y \in \tilE$.

We will then lower-bound this quantity, using Mendelson's small-ball
method (Theorem 8 in \cite{KRT15}):  for any $\xi > 0$ and $t \geq 0$,
with probability at least $1 - e^{-2t^2}$, we have that
\begin{multline}
\label{eqn-porpoise-II}
\inf_{Y \in \tilE} \biggl[ \sum_{i=1}^m \abs{\Tr(Y J(\calC_i) d^2)}^2 \biggr]^{1/2} \\
\geq \xi \sqrt{m} \tilQ_{2\xi}(\tilE) - 2\tilW_m(\tilE) - \xi t, 
\end{multline}
where we define 
\begin{align}
\tilQ_\xi(\tilE) &= \inf_{Y\in \tilE} \left\{\underset{\calC\sim \tilG}{\Pr}[\abs{\Tr(Y J(\calC) d^2)} \geq \xi]\right\}, \\
\tilW_m(\tilE) &= \underset{\substack{\epsilon_i\sim\pm1\\\calC\sim \tilG}}\EE\left[ \sup_{Y\in \tilE} \Tr(Y\tilH)\right], \nonumber\\
& \textrm{with } \tilH = \frac{1}{\sqrt{m}} \sum_{i=1}^m \eps_i J(\calC_i) d^2, 
\end{align}
and $\eps_1,\ldots,\eps_m$ are Rademacher random variables.  

%%%%%%%%%%%%%%%%%%%%%%%%%%%%%%%%%%%%%%%%%%%%%%%%%%%%%%%%

\subsection{Lower-bounding $\tilQ_\xi(\tilE)$}

We will lower-bound $\tilQ_\xi(\tilE)$ as follows.  Fix some $Y \in \tilE$.  We
know that $Y$ is in $\tilE(\calU)$, for some unitary process $\calU:\: \rho \mapsto U\rho U^\dagger$ such
that $U$ is non-spiky in the sense of (\ref{eqn-non-spiky-II}). 
(As a result, $J(\calU)d$ also satisfies the last three constraints in 
(\ref{eqn-silver}).) 
Hence, we know that $\norm{Y}_F = 1$, and there exists some $\tau > 0$
such that $J(\calU)d + \tau Y$ is non-spiky in the sense of 
(\ref{eqn-silver-nonspiky}) and satisfies the last three constraints in 
(\ref{eqn-silver}), and we have
\begin{equation}
\Tr(J(\calU)d + \tau Y) \leq \Tr(J(\calU)d), \quad 
\norm{\tau Y}_F \geq \delta d.  
\end{equation}
Note that we have $\norm{\tau Y}_F = \tau \geq \delta d$.  

We will lower-bound $\Pr[\abs{\Tr(Y J(\calC) d^2)} \geq \xi]$, for any
$\xi \in [0,1]$, using the Paley-Zygmund inequality, and appropriate
bounds on the second and fourth moments of $\Tr(Y J(\calC) d^2)$.  To
simplify the notation, let us define a random variable 
\begin{equation}
\tilS = \tau \abs{\Tr(Y J(\calC) d^2)}.
\end{equation}

%%%%%%%%%%%%%%%%%%%%%%%%%%%%%%%%%%%%%%%%%%%%%%%%%%%%%%%%

\subsubsection{Lower-bounding $\EE(\tilS^2)$}\label{sec:bound_second_moment}

%Let 
%\begin{align}
%S_i=|\tr(J(\sop C_i)(J(\sop Z)-J(\sop U_0)))|.
%\end{align}
%We will use the Payley-Zygmund inequality to bound $S_i^2$:
%\begin{align}\label{eq:PZ1}
%\mathbb P_{\sop C_i\in \Clifs}
%\left[S_i^2\geq \frac{1}{2}\mathbb  
%E_{\sop C_i\in \Clifs}[S_i^2]\right]\geq 
%\frac{E_{\sop C_i\in \Clifs}[S_i^2]^2}{4E_{\sop C_i\in \Clifs}[S_i^4]}.
%\end{align}

We will first put $\tilS$ into a form that is easier to work with.
We can write $Y$ in the form $Y = J(\calY)d$, which is the rescaled Choi matrix of some linear map 
$\calY \in \calL(\CC^{d\times d}, \CC^{d\times d})$.  
Using the relationship between the trace of Choi and Liouville 
representations (see Appendix \ref{app:small_calcs}), we have
\begin{align}
\tilS =& \tau |\tr( J(\sop C) J(\calY) d^3 )| \nonumber\\
=& \tau d |\tr( (\sop C^L)^\dagger \calY^L )| \nonumber\\
=& \tau d |\tr( (C^\dagger \otimes C^T) \calY^L )|.
\end{align}

We can calculate $\EE(\tilS^2)$ by using the fact that $C \in \tilG$ is chosen from a unitary 2-design:
\begin{align}
%\underset{\calC\sim\tilG}{\EE}
\EE(\tilS^2) 
= &\tau^2 d^2 \int_{\textrm{Haar}} \left|\tr\left( (U \otimes U^* \otimes U \otimes U^*) (\calY^L \otimes \calY^L) \right)\right| dU \nonumber\\
\geq &\tau^2 d^2 \left| \tr\left( \int_{\textrm{Haar}} (U \otimes U^* \otimes U \otimes U^*) dU \, (\calY^L \otimes \calY^L) \right) \right|.
\label{eq:second_moment}
\end{align}

Now using Weingarten functions \cite{collins03, collins06, scott2008optimizing}, we have that
\begin{align}\label{eq:simple_wein}
\int_{\textrm{Haar}} (U \otimes& U \otimes U^\dagger \otimes U^\dagger) dU \nonumber\\
&= \frac{P_{3412}+P_{4321}}{d^2-1}-\frac{P_{3421}+P_{4312}}{d(d^2-1)}
\end{align}
where 
\begin{align}
&P_{\sigma(1),\sigma(2),...\sigma(t)} \nonumber\\
&\quad= \sum_{j_1,j_2,\dots,j_t} \ketbra{j_1}{j_{\sigma(1)}} \otimes
\ketbra{j_2}{j_{\sigma(2)}} \otimes \cdots \otimes \ketbra{j_t}{j_{\sigma(t)}}
\end{align}
is a permutation of the registers.

Let $\mathbb T_2(\cdot)$ transpose the second half of the indices of a matrix in $\CC^{d^2\times d^2}$.
That is, for any matrix 
\begin{align}
X = \sum_{ijkl} x_{ijkl} \ketbra{i}{j} \otimes \ketbra{k}{l},
\end{align}
we have 
\begin{align}
\mathbb T_2(X) = \sum_{ijkl} x_{ijkl} \ketbra{i}{j} \otimes \ketbra{l}{k}.
\end{align}
Then note that
\begin{align}
P_{1324} \mathbb T_2 \left( \int_{\textrm{Haar}}
(U \otimes U \otimes U^\dagger \otimes U^\dagger) dU\right) P_{1324} \nonumber\\
= \int_{\textrm{Haar}} (U \otimes U^*\otimes U \otimes U^*) dU.
\end{align}

Combining with Eq. (\ref{eq:second_moment}) and Eq. (\ref{eq:simple_wein}),
we have
\begin{widetext}
\begin{align}
\label{eqn-horsefly}
\EE(\tilS^2) \geq \tau^2 d^2 \left| \tr\left( P_{1324} \mathbb T_2
\left(\frac{P_{3412}+P_{4321}}{d^2-1}-\frac{P_{3421}+P_{4312}}{d(d^2-1)}\right)
P_{1324} (\calY^L \otimes \calY^L\right) \right|.
\end{align}
Because addition commutes with permutation and transposition and trace,
we need to calculate
\begin{align}
\tr\left(P_{1324}\mathbb T_2(P_\sigma)P_{1324} (\calY^L\otimes \calY^L) \right)
\end{align}
for $\sigma \in \{3412,4312,3421,4321\}$.  Letting $\sigma=(abcd)$, we have
\begin{align}
\tr&\left( P_{1324} \mathbb T_2(P_{abcd}) P_{1324} (\calY^L \otimes \calY^L) \right) \nonumber\\
= \tr&\left( P_{1324} \mathbb T_2\left(
\sum_{j_1,j_2,j_3,j_4} \ketbra{j_1}{j_a} \otimes
\ketbra{j_2}{j_b} \otimes \ketbra{j_3}{j_c} \otimes
\ketbra{j_4}{j_d} \right) P_{1324} (\calY^L \otimes \calY^L) \right) \nonumber\\
=\tr&\left( P_{1324}
\sum_{j_1,j_2,j_3,j_4} \ketbra{j_1}{j_a} \otimes
\ketbra{j_2}{j_b} \otimes \ketbra{j_c}{j_3} \otimes
\ketbra{j_d}{j_4} P_{1324} (\calY^L \otimes \calY^L) \right) \nonumber\\
=\tr&\left(
\sum_{j_1,j_2,j_3,j_4} \ketbra{j_1}{j_a} \otimes \ketbra{j_c}{j_3} \otimes
\ketbra{j_2}{j_b} \otimes
\ketbra{j_d}{j_4} (\calY^L \otimes \calY^L) \right) \nonumber\\
=& \sum_{j_1,j_2,j_3,j_4} \bra{j_a j_3} \calY^L \ket{j_1 j_c} \bra{j_b j_4} \calY^L \ket{j_2 j_d}.
\label{eq:twotensor}
\end{align}
\end{widetext}

We know that $Y = J(\calY)d$ satisfies the last three constraints in (\ref{eqn-silver}).  Because of these constraints, we have
\begin{equation}
\begin{split}
\sum_{i_1,i_2,i_3} \bra{i_2 i_3} \calY^L \ket{i_1 i_1}
&= \sum_{i_1,i_2,i_3} d \bra{i_2 i_1} J(\calY) \ket{i_3 i_1} \\
&= \sum_{i_2,i_3} d \bra{i_2} \Tr_2(J(\calY)) \ket{i_3} \\
&= d\tr(J(\calY)),
\label{eq:unit_trace}
\end{split}
\end{equation}
and 
\begin{equation}
\begin{split}
\sum_{i_1,i_2,i_3} \bra{i_1 i_1} \calY^L \ket{i_2 i_3}
&= \sum_{i_1,i_2,i_3} d \bra{i_1 i_2} J(\calY) \ket{i_1 i_3} \\
&= \sum_{i_2,i_3} d \bra{i_2} \Tr_1(J(\calY)) \ket{i_3} \\
&= d\tr(J(\calY)).
\label{eq:tp_trace}
\end{split}
\end{equation}

We now go through the four permutations of Eq. (\ref{eqn-horsefly}):
\begin{itemize}
\item ${\mathbf{P_{3412}}}$. In this case, $a=3$, $b=4$,
$c=1$, and $d=2$.  Plugging into Eq. (\ref{eq:twotensor})
and using Eqs. (\ref{eq:tp_trace}) and (\ref{eq:unit_trace}), we have 
\begin{align}
\tr&\left( P_{1324} \mathbb T_2(P_{3412}) P_{1324} (\calY^L \otimes \calY^L) \right) = d^2 \tr(J(\calY))^2.
\end{align}

\item ${\mathbf{P_{4321}}}$. In this case, $a=4$, $b=3$,
$c=2$, and $d=1$.  Plugging into Eq. (\ref{eq:twotensor}), we have 
\begin{align}
\tr&\left( P_{1324} \mathbb T_2(P_{4321}) P_{1324} (\calY^L \otimes \calY^L) \right) \nonumber\\
= &\sum_{j_1,j_2,j_3,j_4} \bra{j_4 j_3} \calY^L \ket{j_1j_2} \bra{j_3 j_4} \calY^L \ket{j_2j_1} \nonumber\\
= &\sum_{j_1,j_2,j_3,j_4} \bra{j_4 j_3} \calY^L \ket{j_1j_2} \bra{j_4 j_3} (\calY^L)^* \ket{j_1j_2} \nonumber\\
= &\sum_{j_1,j_2,j_3,j_4} |\bra{j_4 j_3} \calY^L \ket{j_1j_2}|^2 \nonumber\\
= &\|\calY^L\|_F^2 \nonumber\\
= &d^2 \|J(\calY)\|_F^2,
\end{align}
where we used Eq. (\ref{eq:conj_relation}).

\item ${\mathbf{P_{3421}}}$. In this case, $a=3$, $b=4$,
$c=2$, and $d=1$.  Plugging into Eq. (\ref{eq:twotensor})
and using Eqs. (\ref{eq:tp_trace}) and (\ref{eq:unit_trace}), we have 
\begin{align}
\tr&\left( P_{1324} \mathbb T_2(P_{3412}) P_{1324} (\calY^L \otimes \calY^L) \right) = d^2\tr(J(\calY))^2.
\end{align}
\item ${\mathbf{P_{4312}}}$. In this case, $a=4$, $b=3$,
$c=1$, and $d=2$.  Plugging into Eq. (\ref{eq:twotensor})
and using Eqs. (\ref{eq:tp_trace}) and (\ref{eq:unit_trace}), we have 
\begin{align}
\tr&\left( P_{1324} \mathbb T_2(P_{3412}) P_{1324} (\calY^L \otimes \calY^L) \right) = d^2\tr(J(\calY))^2.
\end{align}
\end{itemize}

Putting it all together, we have
\begin{align}
\label{eq:finalE2}
\EE(\tilS^2)
&\geq \tau^2 d^4 \left| \frac{\tr(J(\calY))^2+\|J(\calY)\|_F^2}{d^2-1} - 
\frac{2\tr(J(\calY))^2}{d(d^2-1)} \right| \nonumber\\
&\geq \tau^2 d^2 (1-\tfrac{2}{d}) \tr(J(\calY))^2 + \tau^2 d^2 \norm{J(\calY)}_F^2 \nonumber\\
&\geq 0 + \tau^2 d^2 \norm{J(\calY)}_F^2
= \tau^2 \norm{Y}_F^2
= \tau^2.
\end{align}

%%%%%%%%%%%%%%%%%%%%%%%%%%%%%%%%%%%%%%%%%%%%%%%%%%%%%%%%

\subsubsection{Upper-bounding $\EE(\tilS^4)$}

To handle $\EE(\tilS^4)$, we need to use a different argument from that in
\cite{KRT15}, since we are sampling from a unitary 2-design, not
a 4-design.  Our solution is to upper-bound $\tilS$, using the 
non-spikiness properties of $J(\calU)d$ and $J(\calU)d + \tau Y$:
\begin{equation}
\begin{split}
\tilS &= \bigl\lvert -\Tr(J(\calU)d J(\calC)d^2) + \Tr((J(\calU)d + \tau Y) J(\calC)d^2) \bigr\rvert \\
&\leq \beta d.
\end{split}
\end{equation}
This implies an upper-bound on $\EE(\tilS^4)$:  
\begin{equation}
\EE(\tilS^4) \leq \beta^2 d^2 \EE(\tilS^2).
\end{equation}

Putting it all together, and using the Paley-Zygmund inequality, for any $\xi \in [0,1]$, we have:  
\begin{equation}
\begin{split}
%\Pr[|\Tr(Y &J(\calC))| \geq \xi / d^2] \\
%&= \Pr[\tilS^2 \geq \xi^2 \tau^2 / d^4] \\
%&\geq \Pr[\tilS^2 \geq \xi^2\EE(\tilS^2)] \\
%&\geq (1-\xi^2)^2 \frac{\EE(\tilS^2)^2}{\EE(\tilS^4)} \\
%&\geq (1-\xi^2)^2 \frac{\tau^2}{d^4} \frac{d^4}{\beta^2} = (1-\xi^2)^2 \frac{\tau^2}{\beta^2}.\\
\Pr[|\Tr(Y &J(\calC) d^2)| \geq \xi] \\
&= \Pr[\tilS^2 \geq \xi^2 \tau^2] \\
&\geq \Pr[\tilS^2 \geq \xi^2\EE(\tilS^2)] \\
&\geq (1-\xi^2)^2 \frac{\EE(\tilS^2)^2}{\EE(\tilS^4)} \\
&\geq (1-\xi^2)^2 \frac{\tau^2}{\beta^2 d^2}.
\end{split}
\end{equation}
Thus, using the fact that $\tau \geq \delta$, we have that 
\begin{equation}
\label{eqn-Q-II}
\tilQ_{\xi}(\tilE) \geq(1-\xi^2)^2\frac{\delta^2}{\beta^2}.
\end{equation}

%%%%%%%%%%%%%%%%%%%%%%%%%%%%%%%%%%%%%%%%%%%%%%%%%%%%%%%%

\subsection{Upper-bounding $\tilW_m(E)$}
\label{sec:unit_WM}

In this section, we will bound
\begin{align}
\tilW_m(E) &= \EE \sup_{Y\in \tilE} \Tr(Y\tilH), 
\quad \tilH = \frac{1}{\sqrt{m}} \sum_{i=1}^m \eps_i J(\calC_i) d^2, 
\end{align}
where the $\eps_i$ are Rademacher random variables,
and the expectation is taken both over the $\eps_i$ 
and the choice of the unitaries $C_i$. For this bound,
 we will only need the fact
that the $C_i$ are chosen from a unitary $1$-design, rather than a 
unitary $2$-design. 

We start by following the argument in \cite{KRT15}, where 
it is shown that 
\begin{align}
 \EE \sup_{Y\in F_r} \Tr(Y\tilH)\leq \EE 2\sqrt{r}\|\tilH\|
\end{align}
where 
\begin{equation}
F_r = \bigcup_{X_{\true}} F(X_{\true}), 
\end{equation}
and the union is over all $X_{\true} \in \CC^{d\times d}_{\text{Herm}}$ with rank at 
most $r$, and 
\begin{equation}
F(X_{\true}) = \set{Y \in D(\norm{\cdot}_*, X_{\true}, 0) \;|\; \norm{Y}_F = 1}.
\end{equation}
This can be compared with our definition of the set $\tilE$ in (\ref{eqn-E-II}) and (\ref{eqn-EXtrue-II}).

In our case, we have $\tilE = \bigcup_{\calU} \tilE(\calU)$, 
where the union is over all processes $\calU$ whose unnormalized Choi state 
$J(\calU)d \in \CC^{d^2\times d^2}_{\text{Herm}}$ has rank $1$, 
and where $\calU$ satisfies the non-spikiness condition (\ref{eqn-silver-nonspiky}).  
$\tilE(\calU)$ is defined similarly to $F(X_{\true})$, but using the function $f$ 
from (\ref{eqn-f-II}) rather than the trace norm.  
While $f$ involves the trace $\Tr(\cdot)$ instead of the trace norm $\norm{\cdot}_*$, 
because we are considering only positive semidefinite matrices, it can be replaced by the trace norm.
Hence $\tilE \subset F_1$, and so
\begin{align}
 \EE \sup_{Y\in \tilE} \Tr(Y\tilH)\leq  \EE \sup_{Y\in F_1} \Tr(Y\tilH)\leq 2\EE \|\tilH\|.
\end{align}

Now we analyze 
\begin{align}
\EE_{\eps, \calC} \left\| \frac{1}{\sqrt{m}} \sum_{i=1}^m \eps_i J(\calC_i) d^2 \right\|,
\end{align}
using similar tools to what is used in \cite{KRT15}.
Since the $\epsilon_j$'s form a Rademacher sequence and $J(\sop C_i)$
are Hermitian, we can apply the non-commutative 
Khintchine inequality \cite{V12,FR13}:
\begin{align}\label{eq:Khin}
\EE_{\eps, \calC}& \left\| \frac{1}{\sqrt{m}} \sum_{i=1}^m 
\eps_i J(\calC_i) d^2 \right\| \nonumber\\
&\leq \mathbb{E}_\calC
\sqrt{2\ln(2d^2)/m} \left\| \sum_{i=1}^mJ(\sop C_i)^2 \right\|^{1/2} d^2 \nonumber\\
&\leq \sqrt{2\ln(2d^2)/m} \left( \mathbb{E}_\calC \left\|
\sum_{i=1}^m J(\sop C_i) \right\| \right)^{1/2} d^2, 
\end{align}
where in the second line we used Jensen's inequality,
and we used the fact that $J(\sop C_i)$ is a rank one
projector with trace 1, so $J(\sop C_i)^2 = J(\sop C_i)$.

We now apply the matrix Chernoff inequality of Theorem 15 in \cite{KRT15}
to $\mathbb{E}_\calC\left\|
\sum_{i=1}^mJ(\sop C_i)\right\|$.
To apply the theorem, we need to bound
\begin{align}
\left\|\mathbb E_\calC\sum_{i=1}^mJ(\sop C_i)\right\| && \textrm{ and }
&& \|J(\sop C_i)\|.
\end{align}

Since $J(\sop C_i)$ corresponds to a quantum state, its maximum eigenvalue
is 1, so $\|J(\sop C_i)\|=1$. Next, notice
\begin{align}
\mathbb E_\calC \sum_{i=1}^m J(\sop C_i) = \sum_{i=1}^m \mathbb E_\calC J(\sop C_i) = \frac{m}{d^2} I.
\end{align}
Because the $C_i$ are drawn from a unitary
2-design, $\mathbb E_\calC J(\sop C_i)$ is the state that results
when a depolarizing channel is applied to one half
of a maximally entangled state. (Randomly applying operations
from a unitary $1$-design results in the depolarizing channel.)
The resulting state is
the maximally mixed state $I/d^2$ in $\CC^{d^2\times d^2}$. Thus
\begin{align}
\left\|\mathbb E_\calC\sum_{i=1}^mJ(\sop C_i)
\right\|=\frac{m}{d^2}.
\end{align}

Then the Matrix Chernoff Inequality (Theorem 15 in \cite{KRT15}) tells
us that
for any $\gamma>0,$ 
%and $m\geq 2d^2\log d^2:$
\begin{align}
\mathbb E \left\| \sum_{j=1}^m J(\sop C_i) \right\|
&\leq \frac{(e^\gamma-1)m + d^2\log d^2}{d^2 \gamma}.
%&\leq \frac{2(e^\gamma-1)+1}{\sqrt{2}{\gamma}}m
\end{align}
Minimizing $\gamma$ gives
\begin{align}
\mathbb E \left\| \sum_{j=1}^m J(\sop C_i) \right\| \leq c_4 m/d^2
\end{align}
for some numerical constant $c_4>0.$
Plugging this expression into Eq. (\ref{eq:Khin}), we obtain the desired bound:
\begin{equation}
\label{eqn-W-II}
\begin{split}
\tilW_m(\tilE) &\leq c_5 \sqrt{\ln d} \cdot d,
\end{split}
\end{equation}
where $c_5 > 0$ is some numerical constant.

%%%%%%%%%%%%%%%%%%%%%%%%%%%%%%%%%%%%%%%%%%%%%%%%%%%%%%%%

\subsection{Final result}

Combining equations (\ref{eqn-dolphin-II}), (\ref{eqn-porpoise-II}), (\ref{eqn-Q-II}) and (\ref{eqn-W-II}), and setting $\xi = 1/4$, $\nu = \beta/\delta$ and $t = \tfrac{1}{16} \sqrt{m} \nu^{-2}$, we get that:
\begin{equation}
\begin{split}
\inf_\calU &\lambda_{\min}(\tilA; D(\tilf, J(\calU)d, \delta d)) \\
&\geq \frac{1}{4} \sqrt{m} \, \frac{9}{16\nu^2} 
- 2c_5 \, \sqrt{\ln d} \cdot d - \frac{1}{4} \, t \\
&= \frac{1}{8} \sqrt{m} \, \frac{1}{\nu^2} 
- 2c_5 \, \sqrt{\ln d} \cdot d. 
\end{split}
\end{equation}
Now we set $m \geq \bigl( 8\nu^2 c_0 \bigr)^2 \cdot d^2 \ln d$, which implies 
\begin{equation}
\frac{1}{8\nu^2 c_0} \sqrt{m} \geq \sqrt{\ln d} \cdot d, 
\end{equation}
hence 
\begin{equation}
\begin{split}
\inf_\calU &\lambda_{\min}(\tilA; D(\tilf, J(\calU)d, \delta d)) \\
&\geq \frac{\sqrt{m}}{8\nu^2} \, \biggl( 1 - \frac{2c_5}{c_0} \biggr).
\end{split}
\end{equation}
This can be plugged into our approximate recovery bound (\ref{eqn-narwhal-II}).  
This finishes the proof of Theorem \ref{thm-main-II}.  $\square$

%%%%%%%%%%%%%%%%%%%%%%%%%%%%%%%%%%%%%%%%%%%%%%%%%%%%%%%%%%%%%%%%%
%%%%%%%%%%%%%%%%%%%%%%%%%%%%%%%%%%%%%%%%%%%%%%%%%%%%%%%%%%%%%%%%%

\section{Phase Retrieval Using Unitary 4-Designs}\label{sec:design_4}

In this section, we again consider the PhaseLift algorithm for recovering an unknown unitary matrix $U \in \CC^{d\times d}$.  However, we now consider a stronger measurement ensemble:  we let the measurement matrices $C_1,\ldots,C_m$ be chosen from a unitary 4-design $\hatG$ in $\CC^{d\times d}$.  This leads to a stronger recovery guarantee:  PhaseLift achieves \textit{exact} recovery of \textit{all} unitaries $U$.  This claim was presented in Theorem \ref{thm-main-III}.  We show the proof here.

We use the same notation as in the previous section, except that here we call the measurement operator $\hatA$ rather than $\tilA$.  We can write the desired solution of the PhaseLift convex program as $J(\calU)d$, and we can write the measurement operator $\hatA$ as 
\begin{equation}
\hatA(\Gamma) = \Bigl[ \Tr(\Gamma J(\calC_i) d^2) \Bigr]_{i=1}^m.
\end{equation}
We want to recover the solution $J(\calU)d$, up to an additive error of size $\delta \norm{J(\calU)d}_F = \delta d$.

%%%%%%%%%%%%%%%%%%%%%%%%%%%%%%%%%%%%%%%%%%%%%%%%%%%%%%%%

\subsection{Descent cone}

For the case of unitary $4$-designs, we use a similar descent cone to
the one used in \cite{KRT15,T14} --- that is, for $f:\: \CC^{d^2\times d^2}_\textrm{Herm} \rightarrow \RR
\cup \set{\infty}$ a proper convex function, and $X_{\true} \in
\CC^{d^2\times d^2}_\textrm{Herm}$, we use the descent cone
$D(f,X_{\true},0).$
This is the set of all directions, originating at the point
$X_{\true}$, that cause the value of $f$ to decrease.

We define the function $\hatf:\: \CC^{d^2\times d^2}_{\Herm} \rightarrow \RR \cup \set{\infty}$ as follows:
\begin{equation}
\label{eqn-f-III}
\hatf(\Gamma) = \begin{cases}
\Tr(\Gamma) &\text{ if $\Gamma$ satisfies the } \\
              &\text{ last three constraints in (\ref{eqn-silver}), } \\
\infty &\text{ otherwise.}
\end{cases}
\end{equation}
By Prop. 7 in \cite{KRT15}, our recovery bound is: 
\begin{equation}
\label{eqn-narwhal-III}
\norm{\Gamma_{\opt} - J(\calU)d}_F \leq  
\frac{2\eta}{\lambda_{\min}(\hatA; D(\hatf, J(\calU)d, 0))},
\end{equation}
and we want to lower-bound the quantity $\lambda_{\min}$: 
\begin{equation}
\label{eqn-dolphin-III}
\begin{split}
\lambda_{\min}(\hatA; D(\hatf, J(\calU)d, 0)) 
= \inf_{Y \in \hatE(\calU)} \biggl[ \sum_{i=1}^m \abs{\Tr(Y J(\caldes_i) d^2)}^2 \biggr]^{1/2}, 
\end{split}
\end{equation}
where we define 
\begin{equation}
\label{eqn-EXtrue-III}
\hatE(\calU) = \set{Y \in D(\hatf, J(\calU)d, 0) \;|\; \norm{Y}_F = 1}.
\end{equation}

%%%%%%%%%%%%%%%%%%%%%%%%%%%%%%%%%%%%%%%%%%%%%%%%%%%%%%%%

\subsection{Mendelson's small-ball method}

We will use Mendelson's small-ball method to lower-bound
$\lambda_{\min}$.  We will prove a uniform lower-bound that holds for
all $\calU$ simultaneously, i.e., we will lower-bound $\inf_{\calU} \lambda_{\min}$.  We define
\begin{equation}
\label{eqn-E-III}
\hatE = \bigcup_{\calU} \hatE(\calU), 
\end{equation}
where the union runs over all unitary processes $\calU \in
\calL(\CC^{d\times d}, \CC^{d\times d})$ of the form 
$\calU:\: \rho \mapsto U\rho U^\dagger$.  We then take the infimum over
all $Y \in \hatE$.

We will then lower-bound this quantity, using Mendelson's small-ball
method (Theorem 8 in \cite{KRT15}):  for any $\xi > 0$ and $t \geq 0$,
with probability at least $1 - e^{-2t^2}$, we have that
\begin{multline}
\label{eqn-porpoise-III}
\inf_{Y \in \hatE} \biggl[ \sum_{i=1}^m \abs{\Tr(Y J(\caldes_i) d^2)}^2 \biggr]^{1/2} \\
\geq \xi \sqrt{m} \hatQ_{2\xi}(\hatE) - 2\hatW_m(\hatE) - \xi t, 
\end{multline}
where we define 
\begin{align}
\hatQ_\xi(\hatE) &= \inf_{Y\in \hatE} \left\{ \underset{C\sim \hatG}{\Pr}[ \abs{\Tr(Y J(\caldes) d^2)} \geq \xi ] \right\}, \\
\hatW_m(\hatE) &= \underset{\substack{\epsilon_i\sim\pm1\\ C_i\sim\hatG}}{\EE} \left[ \sup_{Y\in \hatE} \Tr(Y\hatH) \right], \quad 
\hatH = \frac{1}{\sqrt{m}} \sum_{i=1}^m \eps_i J(\caldes_i) d^2, 
\end{align}
and $\eps_1,\ldots,\eps_m$ are Rademacher random variables.  

%%%%%%%%%%%%%%%%%%%%%%%%%%%%%%%%%%%%%%%%%%%%%%%%%%%%%%%%

\subsection{Lower-bounding $\hatQ_\xi(\hatE)$}

We will lower-bound $\Pr_{C\sim\hatG}[\abs{\Tr(Y J(\caldes) d^2)} \geq \xi]$, for any
$\xi \in [0,1]$, using the Paley-Zygmund inequality, and appropriate
bounds on the second and fourth moments of $|\Tr(Y J(\caldes) d^2)|$.  To
simplify the notation, let us define a random variable 
\begin{equation}
\hatS = \abs{\Tr(Y J(\caldes) d^2)}.
\end{equation}

We will first put $\hatS$ into a form that is easier to work with.
We can write $Y$ in the form $Y = J(\calY)d$, which is the rescaled Choi matrix of some linear map 
$\calY \in \calL(\CC^{d\times d}, \CC^{d\times d})$.  
Using the relationship between the trace of Choi and Liouville 
representations (see Appendix \ref{app:small_calcs}), we have
\begin{align}
\hatS =&  |\tr( J(\calY)d J(\caldes) d^2 )| \nonumber\\
=& d |\tr( (\caldes^L)^\dagger \calY^L )| \nonumber\\
=& d |\tr( (\desU^\dagger \otimes \desU^T) \calY^L )|.
\end{align}

%%%%%%%%%%%%%%%%%%%%%%%%%%%%%%%%%%%%%%%%%%%%%%%%%%%%%%%%

\subsubsection{Lower-bounding $\EE(\hatS^2)$}

Because we are working with a unitary $4$-design, and
$\hatS^2$ only depends on the second moment of the distribution,
the bound will be the same as for a unitary $2$-design, 
and we can use our bound from Section \ref{sec:bound_second_moment}, 
Eq. (\ref{eq:finalE2}) (with $\tau=1$):
\begin{align}
\label{eq:finalE2-III}
\EE(\hatS^2) &\geq d^2 (1-\tfrac{2}{d}) \tr(J(\calY))^2 + d^2 \norm{J(\calY)}_F^2 \nonumber\\
&\geq \tfrac{1}{2} d^2 \max\left\{ \tr(J(\calY))^2, \norm{J(\calY)}_F^2 \right\},
\end{align}
and also (using Eq. (\ref{eq:finalE2})):
\begin{align}\label{eq:finalE2b-III}
\EE(\hatS^2) &\geq d^2 \norm{J(\calY)}_F^2  = \norm{Y}_F^2 = 1.
\end{align}

\subsubsection{Upper-bounding $\EE(\hatS^4)$}

Now we would like to bound $\mathbb E[\hatS^4]$.
Because we are considering a unitary $4$-design, which has the same
fourth moments as the Haar measure on unitaries, instead of taking
the average $\mathbb E[\hatS^4]$ over the $4$-design, we will take the average over Haar random
unitaries.  

We have
\begin{align}\label{eq:fourth_simplify-III}
 \mathbb E[\hatS^4]
=& d^4 \int_{Haar} dU |\tr\left( (U\otimes U^*) \sop Y^L \right)|^4 \nonumber\\
=& d^4 \int_{Haar} dU \nonumber\\
&\left(\tr\left( (U\otimes U^*) \sop Y^L \right)
\tr\left( (U^*\otimes U) (\sop Y^L)^* \right)\right)^2 \nonumber\\
=& d^4 \int_{Haar} dU \nonumber\\
&\tr\left( \left(U\otimes (U^*)^{\otimes 2}\otimes U\right)^{\otimes 2} \left(\sop Y^L\otimes(\sop Y^L)^*\right)^{\otimes 2} \right).
\end{align}

Using similar tricks as in the case of the second moment,
we have that
\begin{align}\label{eq:integral_ordering-III}
&\int_{Haar} dU \left(U\otimes (U^*)^{\otimes 2}\otimes U\right)^{\otimes 2}\nonumber\\
&=P_{15842673}\mathbb{T}_{2}
\left(\int_{Haar} dU U^{\otimes 4}\otimes (U^\dagger)^{\otimes 4}\right)P_{15842673}.
\end{align}
We will define $P^* = P_{15842673}$.
Then we use Weingarten functions \cite{collins03, collins06, scott2008optimizing} 
to obtain an expression for the integral on the
right hand side of Eq. (\ref{eq:integral_ordering-III})
in terms of permutation operators:
\begin{align}\label{eq:weingarten_complex-III}
\int_{Haar} dU U^{\otimes 4}\otimes (U^\dagger)^{\otimes 4}
=\sum_{\sigma,\tau\in \mathbb S_4}W_g(d,4,\sigma\tau^{-1})
P_{\sigma,\tau},
\end{align}
where $\mathbb S_4$ is the symmetric group of 4 elements, and
\begin{align}
&P_{\sigma,\tau}=\nonumber\\
&P_{\tau(1)+4,\tau(2)+4,\tau(3)+4,\tau(4)+4,\sigma^{-1}(1),
\sigma^{-1}(2),\sigma^{-1}(3),\sigma^{-1}(4)}.
\end{align}

Combining Eqs, (\ref{eq:fourth_simplify-III}),
(\ref{eq:integral_ordering-III}), and (\ref{eq:weingarten_complex-III}),
we have
\begin{align}\label{eq:E4big-III}
\mathbb E[\hatS^4] &= d^4 \sum_{\sigma,\tau\in \mathbb S_4} W_g(d,4,\sigma\tau^{-1}) \times \nonumber\\
&\tr\left(P^*\mathbb{T}_{2}
(P_{\sigma,\tau})P^*\left( \sop Y^L\otimes(\sop Y^L)^*\right)^{\otimes 2}\right).
\end{align}

The permutations $P_{\sigma,\tau}$ 
in the sum will be of the form $P_{wxyzabcd}$ (where $(wxyz)$ is a nonrepeating
sequence of elements in the set $\{5,6,7,8\}$
and $(abcd)$ is a non-repeating sequence of elements in the set $\{1,2,3,4\}$. Each term
in the above sum is therefore of the form:
\begin{widetext}
\begin{align}\label{eq:finalY}
\tr&\left(P^*\mathbb{T}_{2}
\left(P_{wxyzabcd}
\right)P^*\left( \sop Y^L\otimes(\sop Y^L)^*\right)^{\otimes 2}\right)\nonumber\\
&=\tr\left(P^*\mathbb{T}_{2}
\left(\sum_{\substack{j_1,j_2,j_3,j_4\\j_5,j_6,j_7,j_8}}\ketbra{j_1,j_2,j_3,j_4,j_5,j_6,j_7,j_8}
{j_w,j_x,j_y,j_z,j_a,j_b,j_c,j_d}
\right)
P^*\left( \sop Y^L\otimes(\sop Y^L)^*\right)^{\otimes 2}\right)\nonumber\\
&=\tr\left(P^*\sum_{\substack{j_1,j_2,j_3,j_4\\j_5,j_6,j_7,j_8}}
\ketbra{j_1,j_2,j_3,j_4,j_a,j_b,j_c,j_d}{j_w,j_x,j_y,j_z,j_5,j_6,j_7,j_8}
P^*\left( \sop Y^L\otimes(\sop Y^L)^*\right)^{\otimes 2}\right)\nonumber\\
&=\tr\left(\sum_{\substack{j_1,j_2,j_3,j_4\\j_5,j_6,j_7,j_8}}
\ketbra{j_1,j_a,j_d,j_4,j_2,j_b,j_c,j_3}{j_w,j_5,j_8,j_z,j_x,j_6,j_7,j_y}
\left( \sop Y^L\otimes(\sop Y^L)^*\right)^{\otimes 2}\right)\nonumber\\
&=\sum_{\substack{j_1,j_2,j_3,j_4\\j_5,j_6,j_7,j_8}}
\bra{j_w,j_5}\sop Y^L\ket{j_1,j_a}\bra{j_8,j_z}(\sop Y^L)^*
\ket{j_d,j_4}\bra{j_x,j_6}\sop Y^L\ket{j_2,j_b}\bra{j_7,j_y}(\sop Y^L)^*\ket{j_c,j_3}
\nonumber\\
&=\sum_{\substack{j_1,j_2,j_3,j_4\\j_5,j_6,j_7,j_8}}
\bra{j_w,j_5}\sop Y^L\ket{j_1,j_a}\bra{j_z,j_8}\sop Y^L
\ket{j_4,j_d}\bra{j_x,j_6}\sop Y^L\ket{j_2,j_b}\bra{j_y,j_7}\sop Y^L\ket{j_3,j_c}\nonumber\\
&=\sum_{\substack{j_1,j_2,j_3,j_4\\j_5,j_6,j_7,j_8}}
\bra{j_1,j_5}\hat{\sop Y}\ket{j_a,j_w}\bra{j_4,j_8}\hat{\sop Y}
\ket{j_d,j_z}\bra{j_2,j_6}\hat{\sop Y}\ket{j_b,j_x}\bra{j_3,j_7}\hat{\sop Y}\ket{j_c,j_y}
\end{align}
where in the second to last line we have used Eq. (\ref{eq:conj_relation}),
and in the last line, we have reordered the elements of $\sop Y^L$ as
\begin{align}\label{eq:reordering}
\bra{j_r,j_s}\sop Y^L\ket{j_t,j_u}=\bra{j_t,j_s}\hat{\sop Y}\ket{j_u,j_r}.
\end{align}

Now we will use a graphical representation of the matrices $\hat{\sop Y}$ in order to evaluate these
terms:
\begin{align}
\begin{tikzpicture}
\tikzstyle{operator} = [draw,fill=white,minimum size=1.5em] 
\node[operator]  (n1) at (0,0) {$\hat{\sop Y}$};
\node (emptyr) at (1,0) {};
\node (emptyl) at (-1,0) {};
\node (equals) at (2,0) {=};
\node (Ftensor) at (4,0) {$\bra{j_t,j_s}\hat{\sop Y}\ket{j_u,j_r}$};
\path[->,font=\scriptsize] ([yshift= -3pt]n1.east) edge node[below] {$r$}  ([yshift= -3pt]emptyr.west);
\path[->,font=\scriptsize] ([yshift= 3pt]n1.east) edge [dashed] node[above] {$u$}  ([yshift= 3pt]emptyr.west);
\path[->,font=\scriptsize] ([yshift= -3pt]emptyl.east) edge node[below] {$s$}  ([yshift= -3pt]n1.west);
\path[->,font=\scriptsize] ([yshift= 3pt]emptyl.east) edge [dashed] node[above] {$t$}  ([yshift= 3pt]n1.west);
\end{tikzpicture}
\end{align}

\begin{align}
\begin{tikzpicture}
\tikzstyle{operator} = [draw,fill=white,minimum size=1.5em] 
\node[operator]  (n1) at (0,0) {$\hat{\sop Y}$};
\node (emptyr) at (1,0) {};
\node (emptyl) at (-1,0) {};
\node[operator] (n2) at (1,0) {$\hat{\sop Y}$};
\node (emptyrr) at (2,0) {};
\node (equals) at (2.7,0) {=};
\node (Ftensor) at (5,0) {$\bra{j_t,j_s}\hat{\sop Y}\hat{\sop Y}\ket{j_m,j_n}$};
\node (equals) at (7,0) {=};
\node (Ftensor) at (10,0) {$\sum_{u,r}\bra{j_t,j_s}\hat{\sop Y}\ketbra{j_u,j_r}{j_u,j_r}\hat{\sop Y}\ket{j_m,j_n}$};
\path[->,font=\scriptsize] ([yshift= -3pt]n1.east) edge node[below] {$r$}  ([yshift= -3pt]n2.west);
\path[->,font=\scriptsize] ([yshift= 3pt]n1.east) edge [dashed] node[above] {$u$}  ([yshift= 3pt]n2.west);
\path[->,font=\scriptsize] ([yshift= -3pt]emptyl.east) edge node[below] {$s$}  ([yshift= -3pt]n1.west);
\path[->,font=\scriptsize] ([yshift= 3pt]emptyl.east) edge [dashed] node[above] {$t$}  ([yshift= 3pt]n1.west);
\path[->,font=\scriptsize] ([yshift= -3pt]n2.east) edge node[below] {$n$}  ([yshift= -3pt]emptyrr.west);
\path[->,font=\scriptsize] ([yshift= 3pt]n2.east) edge [dashed] node[above] {$m$}  ([yshift= 3pt]emptyrr.west);
\end{tikzpicture}
\end{align}

\begin{align}
\begin{tikzpicture}
\tikzstyle{operator} = [draw,fill=white,minimum size=1.5em] 
\node[operator]  (n1) at (0,0) {$\hat{\sop Y}$};
\node (equals) at (2,0) {$=$};
\node (equals) at (6,0) {$\sum_{j_s,j_t}\bra{j_t,j_s}\hat{\sop Y}\ket{j_t,j_s}=\tr(\hat{\sop Y})$};
\path[->,font=\scriptsize] (n1) edge [in=200,out=-20, looseness=7] node[below] {t} (n1) ;
\path[->,font=\scriptsize] (n1) edge [in=160,out=20,dashed,looseness=7] node[above] {s} (n1) ;
\end{tikzpicture}
\end{align}

Furthermore, because of Eq. (\ref{eq:unit_trace}) and Eq. (\ref{eq:tp_trace}),
\begin{align}
\begin{tikzpicture}
\tikzstyle{operator} = [draw,fill=white,minimum size=1.5em] 
\node (emptyll) at (-1,0) {};
\node[operator]  (n1) at (0,0) {$\hat{\sop Y}$};
\node (empty) at (1,0) {};
\node (equals) at (2,0) {$=$};
\node (emptymid) at (3,0) {};
\node[operator]  (n2) at (4,0) {$\hat{\sop Y}$};
\node (emptyr) at (5,0) {};
\node (equals) at (6,0) {$=$};
\node (emptyrr) at (7,0) {}; 
\node[operator] (n3) at (8,0) {$\hat{\sop Y}$};
\node (emptyrrr) at (9,0) {};
\node (equals) at (10,0) {$=$};
\node (finaltrace) at (12,0) {$d\tr(J(\sop Y)).$};
\path[->,font=\scriptsize] (n1) edge [in=200,out=-20, looseness=7] node[below] {$t$} (n1) ;
\path[->,font=\scriptsize] ([yshift= 3pt]n1.east) edge [dashed] node[above] {$u$}  ([yshift= 3pt]empty.west);
\path[->,font=\scriptsize] ([yshift= 3pt]emptyll.east) edge [dashed] node[above] {$r$}  ([yshift= 3pt]n1.west);
\path[->,font=\scriptsize] ([yshift= -3pt]emptymid.east) edge node[below] {$r$}  ([yshift= -3pt]n2.west);
\path[->,font=\scriptsize] ([yshift= -3pt]n2.east) edge node[below] {$u$}  ([yshift= -3pt]emptyr.west);
\path[->,font=\scriptsize] (n2) edge [in=160,out=20,dashed,looseness=7] node[above] {$s$} (n2) ;
\path[->,font=\scriptsize] (n3) edge [in=200,out=-20, looseness=7] node[below] {t} (n3) ;
\path[->,font=\scriptsize] (n3) edge [in=160,out=20,dashed,looseness=7] node[above] {s} (n3) ;
\end{tikzpicture}
\end{align}
\end{widetext}
We see that a single self-loop on either register forces the other register to also have a self
loop. Because of this simplifying effect,
 we can enumerate all possible configurations of commutative diagrams that
arise from choices of $(wxyzabcd)$ in the last line of Eq. (\ref{eq:finalY}).
We have the following 7 diagrams that can represent the last line of Eq. (\ref{eq:finalY}):

\begin{align*}
\begin{tikzpicture}[baseline={([yshift=-.5ex]current bounding box.center)}]
\tikzstyle{operator} = [draw,fill=white,minimum size=1.5em] 
\node (num) at (-1,3) {$\mathbf{I:}$};
\node[operator]  (n1) at (0,0) {$\hat{\sop Y}$};
\node[operator]  (n2) at (2,0) {$\hat{\sop Y}$};
\node[operator] (n3) at (0,2) {$\hat{\sop Y}$};
\node[operator] (n4) at (2,2) {$\hat{\sop Y}$};
\path[->,font=\scriptsize] (n1) edge [in=200,out=-20, looseness=7] (n1) ;
\path[->,font=\scriptsize] (n1) edge [in=160,out=20,dashed,looseness=7] (n1) ;
\path[->,font=\scriptsize] (n2) edge [in=200,out=-20, looseness=7] (n2) ;
\path[->,font=\scriptsize] (n2) edge [in=160,out=20,dashed,looseness=7] (n2) ;
\path[->,font=\scriptsize] (n3) edge [in=200,out=-20, looseness=7] (n3) ;
\path[->,font=\scriptsize] (n3) edge [in=160,out=20,dashed,looseness=7] (n3) ;
\path[->,font=\scriptsize] (n4) edge [in=200,out=-20, looseness=7] (n4) ;
\path[->,font=\scriptsize] (n4) edge [in=160,out=20,dashed,looseness=7] (n4) ;
\end{tikzpicture}
\end{align*}
\begin{align*}
\begin{tikzpicture}[baseline={([yshift=.5ex]current bounding box.center)}]
\tikzstyle{operator} = [draw,fill=white,minimum size=1.5em] 
\node (num) at (-1,3) {$\mathbf{II:}$};
\node[operator]  (n1) at (0,0) {$\hat{\sop Y}$};
\node[operator]  (n2) at (2,0) {$\hat{\sop Y}$};
\node[operator] (n3) at (0,2) {$\hat{\sop Y}$};
\node[operator] (n4) at (2,2) {$\hat{\sop Y}$};
\path[->,font=\scriptsize] ([yshift= -3pt]n1.east) edge ([yshift= -3pt]n2.west);
\path[->,font=\scriptsize] ([yshift= 3pt]n1.east) edge [dashed]  ([yshift= 3pt]n2.west);
\path[->,font=\scriptsize] ([yshift= -3pt]n2.east) edge [in=160,out=20,looseness=2]  ([yshift= -3pt]n1.west);
\path[->,font=\scriptsize] ([yshift= 3pt]n2.east) edge [in=160,out=20,looseness=2,dashed]  ([yshift= 3pt]n1.west);
\path[->,font=\scriptsize] (n3) edge [in=200,out=-20, looseness=7] (n3) ;
\path[->,font=\scriptsize] (n3) edge [in=160,out=20,dashed,looseness=7] (n3) ;
\path[->,font=\scriptsize] (n4) edge [in=200,out=-20, looseness=7] (n4) ;
\path[->,font=\scriptsize] (n4) edge [in=160,out=20,dashed,looseness=7] (n4) ;
\end{tikzpicture}
\end{align*}

\begin{align*}
\begin{tikzpicture}[baseline={([yshift=.5ex]current bounding box.center)}]
\tikzstyle{operator} = [draw,fill=white,minimum size=1.5em] 
\node (num) at (-1,3) {$\mathbf{III:}$};
\node[operator]  (n1) at (0,0) {$\hat{\sop Y}$};
\node[operator]  (n2) at (2,0) {$\hat{\sop Y}$};
\node[operator] (n3) at (0,2) {$\hat{\sop Y}$};
\node[operator] (n4) at (2,2) {$\hat{\sop Y}$};
\path[->,font=\scriptsize] ([yshift= -3pt]n1.east) edge ([yshift= -3pt]n2.west);
\path[->,font=\scriptsize] ([yshift= 3pt]n1.east) edge [dashed] ([yshift= 3pt]n2.west);
\path[->,font=\scriptsize] ([yshift= -3pt]n2.east) edge [in=0,out=0,looseness=.5]  ([yshift= -3pt]n4.east);
\path[->,font=\scriptsize] ([yshift= 3pt]n2.east) edge [in=0,out=0,dashed,looseness=.5]  ([yshift= 3pt]n4.east);
\path[->,font=\scriptsize] (n3) edge [in=200,out=-20, looseness=7] (n3) ;
\path[->,font=\scriptsize] (n3) edge [in=160,out=20,dashed,looseness=7] (n3) ;
\path[->,font=\scriptsize] ([yshift= -3pt]n4.west) edge [in=180,out=180,looseness=.5]  ([yshift= -3pt]n1.west);
\path[->,font=\scriptsize] ([yshift= 3pt]n4.west) edge [in=180,out=180,dashed, looseness=.5]  ([yshift= 3pt]n1.west);
\end{tikzpicture}
\end{align*}

\begin{align*}
\begin{tikzpicture}[baseline={([yshift=.5ex]current bounding box.center)}]
\tikzstyle{operator} = [draw,fill=white,minimum size=1.5em] 
\node (num) at (-1,3) {$\mathbf{IV:}$};
\node[operator]  (n1) at (0,0) {$\hat{\sop Y}$};
\node[operator]  (n2) at (2,0) {$\hat{\sop Y}$};
\node[operator] (n3) at (0,2) {$\hat{\sop Y}$};
\node[operator] (n4) at (2,2) {$\hat{\sop Y}$};
\path[->,font=\scriptsize] ([yshift= -3pt]n1.east) edge ([yshift= -3pt]n2.west);
\path[->,font=\scriptsize] ([yshift= 3pt]n1.east) edge [dashed] ([yshift= 3pt]n2.west);
\path[->,font=\scriptsize] ([yshift= -3pt]n2.east) edge [in=0,out=0,looseness=.5]  ([yshift= -3pt]n4.east);
\path[->,font=\scriptsize] ([yshift= 3pt]n2.east) edge [in=0,out=0,dashed,looseness=.5]  ([yshift= 3pt]n4.east);
\path[->,font=\scriptsize] ([yshift= -3pt]n3.west) edge[in=180,out=180,looseness=.5] ([yshift= -3pt]n1.west);
\path[->,font=\scriptsize] ([yshift= 3pt]n3.west) edge [in=180,out=180,looseness=.5,dashed] ([yshift= 3pt]n1.west);
\path[->,font=\scriptsize] ([yshift= -3pt]n4.west) edge [in=0,out=180,looseness=.5]  ([yshift= -3pt]n3.east);
\path[->,font=\scriptsize] ([yshift= 3pt]n4.west) edge [dashed]  ([yshift= 3pt]n3.east);
\end{tikzpicture}
\end{align*}

\begin{align}
\begin{tikzpicture}[baseline={([yshift=.5ex]current bounding box.center)}]
\tikzstyle{operator} = [draw,fill=white,minimum size=1.5em] 
\node (num) at (-1,3) {$\mathbf{V:}$};
\node[operator] (n1) at (0,0) {$\hat{\sop Y}$};
\node[operator] (n2) at (2,0) {$\hat{\sop Y}$};
\node[operator] (n3) at (0,2) {$\hat{\sop Y}$};
\node[operator] (n4) at (2,2) {$\hat{\sop Y}$};
\path[->,font=\scriptsize] ([yshift= -3pt]n1.east) edge [in=0,out=0,looseness=.5] ([yshift= -3pt]n3.east);
\path[->,font=\scriptsize] ([yshift= 3pt]n1.east) edge [dashed] ([yshift= 3pt]n2.west);
\path[->,font=\scriptsize] ([yshift= -3pt]n2.east) edge [in=0,out=0,looseness=.5]  ([yshift= -3pt]n4.east);
\path[->,font=\scriptsize] ([yshift= 3pt]n2.east) edge [in=0,out=0,dashed,looseness=.5]  ([yshift= 3pt]n4.east);
\path[->,font=\scriptsize] ([yshift= -3pt]n3.west) edge[in=180,out=180,looseness=.5] ([yshift= -3pt]n1.west);
\path[->,font=\scriptsize] ([yshift= 3pt]n3.west) edge [in=180,out=180,looseness=.5,dashed] ([yshift= 3pt]n1.west);
\path[->,font=\scriptsize] ([yshift= -3pt]n4.west) edge [in=180,out=180,looseness=.5]  ([yshift= -3pt]n2.west);
\path[->,font=\scriptsize] ([yshift= 3pt]n4.west) edge [dashed]  ([yshift= 3pt]n3.east);
\end{tikzpicture}
\end{align}

\begin{align}
\begin{tikzpicture}[baseline={([yshift=.5ex]current bounding box.center)}]
\tikzstyle{operator} = [draw,fill=white,minimum size=1.5em] 
\node (num) at (-1,3) {$\mathbf{VI:}$};
\node[operator] (n1) at (0,0) {$\hat{\sop Y}$};
\node[operator] (n2) at (2,0) {$\hat{\sop Y}$};
\node[operator] (n3) at (0,2) {$\hat{\sop Y}$};
\node[operator] (n4) at (2,2) {$\hat{\sop Y}$};
\path[->,font=\scriptsize] ([yshift= -3pt]n1.east) edge [in=0,out=0,looseness=.5] ([yshift= -3pt]n3.east);
\path[->,font=\scriptsize] ([yshift= -3pt]n2.east) edge [in=0,out=0,looseness=.5]  ([yshift= -3pt]n4.east);
\path[->,font=\scriptsize] ([yshift= -3pt]n3.west) edge[in=180,out=180,looseness=.5] ([yshift= -3pt]n1.west);
\path[->,font=\scriptsize] ([yshift= -3pt]n4.west) edge [in=180,out=180,looseness=.5]  ([yshift= -3pt]n2.west);
\path[->,font=\scriptsize] ([yshift= 3pt]n1.east) edge [dashed] ([yshift= 3pt]n2.west);
\path[->,font=\scriptsize] ([yshift= 3pt]n2.east) edge [in=160,out=20,dashed,looseness=2]  ([yshift= 3pt]n1.west);
\path[->,font=\scriptsize] ([yshift= 3pt]n3.west) edge [in=20,out=160,looseness=2,dashed] ([yshift= 3pt]n4.east);
\path[->,font=\scriptsize] ([yshift= 3pt]n4.west) edge [dashed]  ([yshift= 3pt]n3.east);
\end{tikzpicture}
\end{align}

\begin{align}
\begin{tikzpicture}[baseline={([yshift=.5ex]current bounding box.center)}]
\tikzstyle{operator} = [draw,fill=white,minimum size=1.5em] 
\node (num) at (-1,3) {$\mathbf{VII:}$};
\node[operator] (n1) at (0,0) {$\hat{\sop Y}$};
\node[operator] (n2) at (2,0) {$\hat{\sop Y}$};
\node[operator] (n3) at (0,2) {$\hat{\sop Y}$};
\node[operator] (n4) at (2,2) {$\hat{\sop Y}$};
\path[->,font=\scriptsize] ([yshift= -3pt]n1.east) edge [in=0,out=0,looseness=.5] ([yshift= -3pt]n3.east);
\path[->,font=\scriptsize] ([yshift= 3pt]n1.east) edge [in=0,out=0,dashed, looseness=.5] ([yshift= 3pt]n3.east);
\path[->,font=\scriptsize] ([yshift= -3pt]n2.east) edge [in=0,out=0,looseness=.5]  ([yshift= -3pt]n4.east);
\path[->,font=\scriptsize] ([yshift= 3pt]n2.east) edge [in=0,out=0,dashed,looseness=.5]  ([yshift= 3pt]n4.east);
\path[->,font=\scriptsize] ([yshift= -3pt]n3.west) edge[in=180,out=180,looseness=.5] ([yshift= -3pt]n1.west);
\path[->,font=\scriptsize] ([yshift= 3pt]n3.west) edge[in=180,out=180,dashed,looseness=.5] ([yshift= 3pt]n1.west);
\path[->,font=\scriptsize] ([yshift= -3pt]n4.west) edge [in=180,out=180,looseness=.5]  ([yshift= -3pt]n2.west);
\path[->,font=\scriptsize] ([yshift= 3pt]n4.west) edge [in=180,out=180,dashed,looseness=.5]  ([yshift= 3pt]n2.west);
\end{tikzpicture}
\end{align}

These commutative diagrams were found in the following way. First, there is the case that all four $\hat{\sop Y}$'s have self loops. This corresponds to Diagram I. Next, we consider the case that three $\hat{\sop Y}$'s have self loops, but in this case, the only way to connect up the final $\hat{\sop Y}$'s tensor legs is to create self loops, so we are back to Diagram I. In the case that two $\hat{\sop Y}$'s have self loops, the only way to connect the remaining two $\hat{\sop Y}$'s without giving them self loops is as shown in Diagram II. Continuing in this way, we can enumerate all possible diagrams.

We ignore diagrams that are identical to diagrams that are depicted, but which have
one or more loops reversed in direction, or that have dashed and solid arrows switched. This is acceptable, because by reordering the elements of the matrix,
as we did in Eq. (\ref{eq:reordering}), we can create a new figure which
looks identical to ones shown above, but involving a new matrix $\hat{\sop Y}'$ whose
elements are the same as $\hat{\sop Y}$. In our analysis below, we will ultimately
see that our bounds on the contributions due to each figure will depend only on
the Frobenius norm of $\sop Y^L$, which only depends on the elements of the matrix,
and not on their ordering. (We also have contributions that depend on the $\tr(\sop Y)$,
but these terms only come from self-loops about a single element, for which reordering
does not produce a new term.)

For a square matrix $A$ and integer $i>1$, we will use the following bound:
\begin{align}
\tr(A^i)&=\tr(A^{i-1}A)\nonumber\\
&=\textrm{vec}((A^{i-1})^T)^T\textrm{vec}(A)\nonumber\\
&\leq\|(A^{i-1})^T\|_F\|A\|_F\nonumber\\
&\leq\|A\|^i_F
\end{align}
where we have used Cauchy-Schwarz, the submultiplicative property of the Frobenius
norm, and the fact that $\|A^T\|_F=\|A\|_F$.

We now bound the contribution due to each diagram.
\begin{itemize}
\item[\bf{I:}] We read off the contribution of
$\tr(\hat{\sop Y})^4=d^4\tr(J(\sop Y))^4$.
\item[\bf{II:}] We have a contribution of
\begin{align}
d^2\tr(J(\sop Y))^2\tr(\hat{\sop Y}^2)\leq &d^2\tr(J(\sop Y))^2\|{\hat{\sop Y}}\|_F^2\nonumber\\
=&d^4\tr(J(\sop Y))^2\|J(\sop Y)\|_F^2.
\end{align}
\item[\bf{III:}] We have a contribution of
\begin{align}
d\tr(J(\sop Y))\tr(\hat{\sop Y}^3)\leq &d\tr(J(\sop Y))\|\hat{\sop Y}^2\|_F^3\nonumber\\
=&d^4\tr(J(\sop Y))\|J(\sop Y)\|_F^3.
\end{align}
\item[\bf{IV:}] We have a contribution of
\begin{align}
\tr(\hat{\sop F}^4)\leq d^4\|J(\sop F)\|_F^4.
\end{align}
\item[\bf{V:}] We reprint the diagram here, but with labels:
\begin{align}
\framebox{
\begin{tikzpicture}[baseline={([yshift=.5ex]current bounding box.center)}]
\tikzstyle{operator} = [draw,fill=white,minimum size=1.5em] 
\node (num) at (-1,3) {$\mathbf{V:}$};
\node[operator] (n1) at (0,0) {$\hat{\sop Y}$};
\node[operator] (n2) at (2,0) {$\hat{\sop Y}$};
\node[operator] (n3) at (0,2) {$\hat{\sop Y}$};
\node[operator] (n4) at (2,2) {$\hat{\sop Y}$};
\path[->,font=\scriptsize] ([yshift= -3pt]n1.east) edge [in=0,out=0,looseness=.5] node[left] {$\beta$} ([yshift= -3pt]n3.east);
\path[->,font=\scriptsize] ([yshift= 3pt]n1.east) edge [dashed] node[below] {$\xi$} ([yshift= 3pt]n2.west);
\path[->,font=\scriptsize] ([yshift= -3pt]n2.east) edge [in=0,out=0,looseness=.5] node[left] {$\gamma$} ([yshift= -3pt]n4.east);
\path[->,font=\scriptsize] ([yshift= 3pt]n2.east) edge [in=0,out=0,dashed,looseness=1] node[right] {$\psi$}  ([yshift= 3pt]n4.east);
\path[->,font=\scriptsize] ([yshift= -3pt]n3.west) edge[in=180,out=180,looseness=.5] node[right] {$\alpha$} ([yshift= -3pt]n1.west);
\path[->,font=\scriptsize] ([yshift= 3pt]n3.west) edge [in=180,out=180,looseness=1,dashed] node[left] {$\chi$} ([yshift= 3pt]n1.west);
\path[->,font=\scriptsize] ([yshift= -3pt]n4.west) edge [in=180,out=180,looseness=.5]  node[left] {$\delta$} ([yshift= -3pt]n2.west);
\path[->,font=\scriptsize] ([yshift= 3pt]n4.west) edge [dashed] node[above] {$\phi$} ([yshift= 3pt]n3.east);
\end{tikzpicture}}
\end{align}

Let $K$ be the $d\times d$ matrix such that
\begin{align}
\bra{j_a}K\ket{j_b}=\sqrt{\sum_{j_c,j_d}\left(\bra{j_a,j_c}\hat{\sop Y}\ket{j_b,j_d}\right)^2}.
\end{align}
Then the contribution of the diagram is bounded by
\begin{align}
\sum_{\substack{j_\alpha,j_\beta,j_\gamma,j_\delta\\j_\xi,j_\chi,j_\phi,j_\psi}}
&\bra{j_\phi,j_\beta}\hat{\sop Y}\ket{j_\chi,j_\alpha}\bra{j_\chi,j_\alpha}\hat{\sop Y}
\ket{j_\xi,j_\beta}\nonumber\\
&\times\bra{j_\xi,j_\delta}\hat{\sop Y}\ket{j_\psi,j_\gamma}\bra{j_\psi,j_\gamma}\hat{\sop Y}\ket{j_\phi,j_\delta}\nonumber\\
%=\sum_{\substack{j_\xi,j_\chi\\j_\phi,j_\psi}}&
%\left(\sum_{j_\alpha,j_\beta}\bra{j_\phi,j_\beta}\hat{\sop Y}\ket{j_\chi,j_\alpha}\bra{j_\chi,j_\alpha}\hat{\sop Y}
%\ket{j_\xi,j_\beta}\right)\nonumber\\
%&\times\left(\sum_{j_\gamma,j_\delta}
%\bra{j_\xi,j_\delta}\hat{\sop Y}\ket{j_\psi,j_\gamma}
%\bra{j_\psi,j_\gamma}\hat{\sop Y}\ket{j_\phi,j_\delta}\right)\nonumber\\
\leq\sum_{\substack{j_\xi,j_\chi\\j_\phi,j_\psi}}&
\sqrt{\sum_{j_\alpha,j_\beta}\left(\bra{j_\phi,j_\beta}\hat{\sop Y}\ket{j_\chi,j_\alpha}\right)^2}\nonumber\\
&\times
\sqrt{\sum_{j_\alpha,j_\beta}\left(\bra{j_\chi,j_\alpha}\hat{\sop Y}
\ket{j_\xi,j_\beta}\right)^2}\nonumber\\
&\times\sqrt{
\sum_{j_\gamma,j_\delta}\left(
\bra{j_\xi,j_\delta}\hat{\sop Y}\ket{j_\psi,j_\gamma}\right)^2}\nonumber\\
&\times
\sqrt{\sum_{j_\gamma,j_\delta}\left(\bra{j_\psi,j_\gamma}\hat{\sop Y}\ket{j_\phi,j_\delta}\right)^2}\nonumber\\
\leq\sum_{\substack{j_\xi,j_\chi\\j_\phi,j_\psi}}&\bra{j_\phi}K\ket{j_\chi}
\bra{j_\chi}K\ket{j_\xi}\bra{j_\xi}K\ket{j_\psi}\bra{j_\psi}K\ket{j_\phi}\nonumber\\
&=\tr(K^4)\nonumber\\
&\leq\|K\|_F^4\nonumber\\
&=\left(\sum_{j_a,j_b}\sum_{j_c,j_d}\left(\bra{j_a,j_c}\hat{\sop Y}\ket{j_b,j_d}\right)^2\right)^2\nonumber\\
&=\|\hat{\sop Y}\|_F^4\nonumber\\
&=d^4\|J(\sop Y)\|_F^4.
\end{align}

\item[\bf{VI:}] We reprint the diagram here, but with labels:
\begin{align}
\framebox{
\begin{tikzpicture}[baseline={([yshift=.5ex]current bounding box.center)}]
\tikzstyle{operator} = [draw,fill=white,minimum size=1.5em] 
\node (num) at (-1,3) {$\mathbf{VI:}$};
\node[operator] (n1) at (0,0) {$\hat{\sop Y}$};
\node[operator] (n2) at (2,0) {$\hat{\sop Y}$};
\node[operator] (n3) at (0,2) {$\hat{\sop Y}$};
\node[operator] (n4) at (2,2) {$\hat{\sop Y}$};
\path[->,font=\scriptsize] ([yshift= -3pt]n1.east) edge [in=0,out=0,looseness=.5] node[left] {$\beta$} ([yshift= -3pt]n3.east);
\path[->,font=\scriptsize] ([yshift= -3pt]n2.east) edge [in=0,out=0,looseness=.5] node[left] {$\gamma$} ([yshift= -3pt]n4.east);
\path[->,font=\scriptsize] ([yshift= -3pt]n3.west) edge[in=180,out=180,looseness=.5] node[right] {$\alpha$} ([yshift= -3pt]n1.west);
\path[->,font=\scriptsize] ([yshift= -3pt]n4.west) edge [in=180,out=180,looseness=.5]  node[right] {$\delta$} ([yshift= -3pt]n2.west);
\path[->,font=\scriptsize] ([yshift= 3pt]n1.east) edge [dashed] node[below] {$\psi$} ([yshift= 3pt]n2.west);
\path[->,font=\scriptsize] ([yshift= 3pt]n2.east) edge [in=160,out=20,dashed,looseness=2] node[above] {$\phi$}  ([yshift= 3pt]n1.west);
\path[->,font=\scriptsize] ([yshift= 3pt]n3.west) edge [in=20,out=160,looseness=2,dashed] node[above] {$\chi$} ([yshift= 3pt]n4.east);
\path[->,font=\scriptsize] ([yshift= 3pt]n4.west) edge [dashed] node[below] {$\xi$} ([yshift= 3pt]n3.east);
\end{tikzpicture}}
\end{align}
Let $K'$ be the $d\times d$ matrix such that
\begin{align}
\bra{j_a}K'\ket{j_b}=\sqrt{\sum_{j_c,j_d}\left(\bra{j_a,j_c}\hat{\sop Y}\ket{j_b,j_d}\right)^2}.
\end{align}
Then the contribution of the diagram is bounded by
\begin{align}
\sum_{\substack{j_\alpha,j_\beta,j_\gamma,j_\delta\\j_\xi,j_\chi,j_\phi,j_\psi}}&
\bra{j_\xi,j_\beta}\hat{\sop Y}\ket{j_\chi,j_\alpha}\bra{j_\phi,j_\alpha}\hat{\sop Y}
\ket{j_\psi,j_\beta}\nonumber\\
&\times\bra{j_\psi,j_\delta}\hat{\sop Y}\ket{j_\phi,j_\gamma}\bra{j_\chi,j_\gamma}\hat{\sop Y}\ket{j_\xi,j_\delta}\nonumber\\
%=\sum_{\substack{j_\xi,j_\chi\\j_\phi,j_\psi}}&
%\left(\sum_{j_\alpha,j_\beta}
%\bra{j_\xi,j_\beta}\hat{\sop Y}\ket{j_\chi,j_\alpha}\bra{j_\phi,j_\alpha}\hat{\sop Y}
%\ket{j_\psi,j_\beta}\right)\nonumber\\
%&\times
%\left(\sum_{j_\gamma,j_\delta}\bra{j_\psi,j_\delta}\hat{\sop Y}\ket{j_\phi,j_\gamma}\bra{j_\chi,j_\gamma}\hat{\sop Y}\ket{j_\xi,j_\delta}\right)\nonumber\\
\leq\sum_{\substack{j_\xi,j_\chi\\j_\phi,j_\psi}}&
\sqrt{\sum_{j_\alpha,j_\beta}\left(\bra{j_\xi,j_\beta}\hat{\sop Y}\ket{j_\chi,j_\alpha}\right)^2}\nonumber\\
&\times
\sqrt{\sum_{j_\alpha,j_\beta}\left(\bra{j_\phi,j_\alpha}\hat{\sop Y}
\ket{j_\psi,j_\beta}\right)^2}\nonumber\\
&\times\sqrt{
\sum_{j_\gamma,j_\delta}\left(\bra{j_\psi,j_\delta}\hat{\sop Y}\ket{j_\phi,j_\gamma}\right)^2}\nonumber\\
&\times
\sqrt{\sum_{j_\gamma,j_\delta}\left(\bra{j_\chi,j_\gamma}\hat{\sop Y}\ket{j_\xi,j_\delta}\right)^2}\nonumber\\
\leq\sum_{\substack{j_\xi,j_\chi\\j_\phi,j_\psi}}&\bra{j_\xi}K'\ket{j_\chi}
\bra{j_\chi}K'\ket{j_\xi}\bra{j_\phi}K'\ket{j_\psi}\bra{j_\psi}K'\ket{j_\phi}\nonumber\\
&=\tr(K'^2)^2\nonumber\\
&\leq\|K'\|_F^4\nonumber\\
&=\left(\sum_{j_a,j_b}\sum_{j_c,j_d}\left(\bra{j_a,j_c}\hat{\sop Y}\ket{j_b,j_d}\right)^2\right)^2\nonumber\\
&=\|\hat{\sop Y}\|_F^4\nonumber\\
&=d^4\|J(\sop Y)\|_F^4.
\end{align}
\item[\bf{VII:}] We have a contribution of
\begin{align}
\tr(\hat{\sop Y}^2)\tr(\hat{\sop Y}^2)\leq &\|\hat{\sop Y}\|_F^4\nonumber\\
\leq& d^4\|J(\sop Y)\|_F^4.
\end{align}
\end{itemize}

Looking at all possible diagrams, we see that the total contribution of any diagram is
less than
\begin{align}
d^4\max\{\tr(J(\sop Y))^4,\|J(\sop Y)\|_F^4\}.
\end{align}
Hence this bounds the size of any term in Eq. (\ref{eq:E4big-III}). Each 
term in Eq. (\ref{eq:E4big-III}) is multiplied by a Weingarten function
$W_g(d,4,\sigma\tau^{-1})$.
The largest Weingarten term is
\begin{align}
&W_g(d,4,(1234))=\nonumber\\
&\frac{d^4-8d^2+6}{(d+3)(d+2)(d+1)(d-1)(d-2)(d-3)d^2}\nonumber\\
<&\frac{2}{d^{4}}
\end{align}
for $d>3$.
There are $(4!)^2$ total terms in the sum. Putting it all together, we have
\begin{align}\label{eq:finalE4}
\mathbb E[\hatS^4] \leq 2(4!)^2 d^4 \max\{ \tr(J(\sop Y))^4, \|J(\sop Y)\|_F^4 \}
\end{align}
for $d>3.$

Now we combine Eqs. (\ref{eq:finalE2-III}), (\ref{eq:finalE2b-III}), and (\ref{eq:finalE4}), and the Payley-Zygmund
inquality to obtain
\begin{align}
\mathbb P\left[ \hatS \geq \xi \right]
&\geq\mathbb P\left[ \hatS^2 \geq \xi^2\mathbb{E}[\hatS^2] \right] \nonumber\\
&\geq \frac{(1-\xi^2)^2 (\mathbb E \hatS^2)^2}{\mathbb E \hatS^4} \nonumber\\
&\geq \frac{(1-\xi^2)^2 \frac{1}{4} d^4 \max\{ \tr(J(\sop F))^4, \|J(\sop F)\|_F^4 \}
}{2(4!)^2 d^4 \max\{ \tr(J(\sop F))^4, \|J(\sop F)\|_F^4 \}} \nonumber\\
&\geq \frac{(1-\xi^2)^2}{8(4!)^2}.
\end{align}

Thus
\begin{equation}
\label{eqn-Q-III}
\hatQ_{\xi}(\hatE) \geq \frac{(1-\xi^2)^2}{8(4!)^2}.
\end{equation}

\subsection{Upper-bounding $\hatW_m(\hatE)$}
\label{sec:unit_WM-III}

In this section, we will bound
\begin{align}
\hatW_m(\hatE) &= \EE \sup_{Y\in \hatE} \Tr(Y\hatH), 
\quad \hatH = \frac{1}{\sqrt{m}} \sum_{i=1}^m \eps_i J(\caldes_i) d^2, 
\end{align}
where the $\eps_i$ are Rademacher random variables,
and the expectation is taken both over the $\eps_i$ 
and the choice of the unitaries $\desU_i$ in the 
unitary $4$-design. Because a unitary $4$-design is also 
a unitary $2$-design, a nearly identical argument 
to that used in Sec. \ref{sec:unit_WM}
holds, and we have
\begin{equation}
\label{eqn-W-III}
\begin{split}
\hatW_m(\hatE) &\leq c_5 \sqrt{\ln d} \cdot d,
\end{split}
\end{equation}
where $c_5 > 0$ is some numerical constant.

\subsection{Final result}

Combining equations (\ref{eqn-dolphin-III}), (\ref{eqn-porpoise-III}),
(\ref{eqn-Q-III}) and (\ref{eqn-W-III}), and setting $\xi = 1/4$
and $t = \tfrac{1}{16} \cdot \tfrac{1}{8(4!)^2} \sqrt{m}$, we get that:
\begin{equation}
\begin{split}
\inf_{\calU} &\lambda_{\min}(\hatA; D(\hatf, J(\calU)d, 0)) \\
&\geq \tfrac{1}{4} \sqrt{m} \tfrac{9}{16} \cdot \tfrac{1}{8(4!)^2} 
- 2c_5 \, \sqrt{\ln d} \cdot d - \tfrac{1}{4} \, t \\
&= \tfrac{1}{4} \sqrt{m} \tfrac{1}{2} \cdot \tfrac{1}{8(4!)^2} 
- 2c_5 \, \sqrt{\ln d} \cdot d.
\end{split}
\end{equation}
Now we set $m \geq \bigl( 64(4!)^2 c_0 \bigr)^2 \cdot d^2 \ln d$, which implies 
\begin{equation}
\frac{1}{64(4!)^2 c_0} \sqrt{m} \geq \sqrt{\ln d} \cdot d, 
\end{equation}
hence 
\begin{equation}
\begin{split}
\inf_{\calU} &\lambda_{\min}(\hatA; D(\hatf, J(\calU)d, 0)) \\
&\geq \frac{\sqrt{m}}{64 (4!)^2} \biggl(1 - \frac{2c_5}{c_0}\biggr).
\end{split}
\end{equation}
This can be plugged into our approximate recovery bound (\ref{eqn-narwhal-III}).  
This finishes the proof of Theorem \ref{thm-main-III}.  $\square$

%%%%%%%%%%%%%%%%%%%%%%%%%%%%%%%%%%%%%%%%%%%%%%%%%%%%%

\section{Quantum Process Tomography}
\label{sec:tomog}

\subsection{Motivation}

Quantum process tomography is an important tool for the experimental
development of large-scale quantum information processors.  Process
tomography is a means of obtaining complete knowledge of the dynamical
evolution of a quantum system.  This allows for accurate calibration
of quantum gates, as well as characterization of qubit noise
processes, such as dephasing, leakage, and cross-talk.  These are
the types of error processes that must be understood and ameliorated
in order build scalable quantum computers with error rates below
the fault-tolerance threshold.

Formally, a quantum process $\calE$ is described by a completely-positive trace-preserving linear map $\calE:\: \CC^{d\times d} \rightarrow \CC^{d\times d}$, which maps density matrices to density matrices.  We want to learn $\calE$ by applying it on known input states $\rho$, and measuring the output states $\calE(\rho)$.  

Process tomography is challenging for (at least) two reasons.  First,
it requires estimating a \textit{large} number of parameters:  for a
system of $n$ qubits, the Hilbert space has dimension $d=2^n$, and a
general quantum process has approximately $d^4 = 2^{4n}$ degrees of freedom.  For
example, to characterize the cross-talk between a pair of two-qubit
gates acting on $n=4$ qubits, using the most general approach, one
must estimate $\sim2^{16}$ parameters.  To address this issue, several
authors have proposed methods based on \textit{compressed sensing},
which exploit sparse or low-rank structure in the unknown state or
process, to reduce the number of measurements that must be performed
\cite{gross10, cramer10, flammia12, shabani11, shabani11b, baldwin14,
kalev15}.

The second challenge is that, in most real experimental setups, one
must perform process tomography using state preparation and
measurement devices that are \textit{imperfect}.  These devices
introduce state preparation and measurement errors (``SPAM errors''),
which limit the accuracy of process tomography.  This limit is
encountered in practice, and often produces non-physical estimates of
processes \cite{JSR15}.  Perhaps
surprisingly, there are methods that are robust to SPAM errors, such
as \textit{randomized benchmarking} \cite{knill08, magesan11,
magesan12, K14, JSR15}, and gate set tomography \cite{merkel13, stark14,
greenbaum15}.

Here, we show how our phase retrieval techniques can address both of these challenges.  
We focus on the special case where we want to learn a \textit{unitary} quantum process.  
This is a process $\calU:\: \CC^{d\times d} \rightarrow \CC^{d\times d}$ that has the form 
\begin{equation}
\label{eqn-whelk}
\calU:\: \rho \mapsto U\rho U^\dagger, 
\end{equation}
where $U \in \CC^{d\times d}$ is a unitary matrix.  This describes the dynamics of a closed quantum system, e.g., time evolution generated by some Hamiltonian, or the action of unitary gates in a quantum computer.  
Using our phase retrieval techniques, we devise methods for learning $\calU$ that are 
both fast \textit{and} robust to SPAM errors (at least in principle).  

Whereas a generic quantum process would have $\sim d^4$ degrees of freedom, $\calU$ has only $\sim d^2$ degrees of freedom.  Our phase retrieval techniques work by estimating $m = O(d^2 \text{poly}(\log d))$ parameters of $\calU$, hence they achieve a quadratic speedup over conventional tomography.  Furthermore, we show how these measurements can be implemented using commonly-available quantum operations.  In particular, they can be performed using randomized benchmarking protocols, which are robust to SPAM errors.

%%%%%%%%%%%%%%%%%%%%%%%%%%%%%%%%%%%%%%%%%%%%%%

\subsection{Measurement Techniques}

In order to learn the unknown process $\calU$, we would like to measure the quantities $\abs{\Tr(C_i^\dagger U)}^2$, where $C_1,\ldots,C_m$ are chosen at random from a unitary 4- or 2-design in $\CC^{d\times d}$.  Then we can apply the PhaseLift algorithm to reconstruct $U$.  

We remark that these measurements are quite different from the Pauli measurements that were used in earlier works on compressed sensing for quantum tomography \cite{gross10, liu11, flammia12}.  In those earlier works, one would estimate the Pauli expectation values of the Jamiolkowski state $J(\calU)$.  This led to measurements of the form $\Tr((P_1 \tensr P_2) J(\calU)) = \frac{1}{d} \Tr(P_1 U P_2 U^\dagger)$, where $P_1$ and $P_2$ were multi-qubit Pauli operators.

There are (at least) two ways of measuring the quantity $\abs{\Tr(C^\dagger U)}^2$, for some prescribed unitary $C \in \CC^{d\times d}$.  The first way works by estimating the Choi-Jamiolkowski state of $U$.  This method is straightforward, and fairly efficient, but it requires reliable and error-free state preparations and measurements.  

To apply this method, recall that any quantum process $\calE$ can be equivalently described by its Choi-Jamiolkowski state, 
\begin{equation}
J(\calE) = (\calE \tensr \calI) (\ket{\Phi^+}\bra{\Phi^+}) \in \CC^{d^2\times d^2}, 
\end{equation}
which is obtained by applying $\calE$ to one half of the maximally entangled state 
%\begin{equation}
$\ket{\Phi^+} = \frac{1}{\sqrt{d}} \sum_{i=0}^{d-1} \ket{i} \tensr \ket{i} \in \CC^{d^2}$.
%\end{equation}
In the case of a unitary process $\calU$, this is equivalent to the ``lifted'' representation of $U$ used in PhaseLift:
\begin{equation}
\label{eqn-J-U}
\begin{split}
J(\calU) &= (U\tensr I) \ket{\Phi^+} \bra{\Phi^+} (U^\dagger \tensr I) \\
&= \frac{1}{d} \VEC(U)\VEC(U)^\dagger.
\end{split}
\end{equation}
In particular, this implies that 
\begin{equation}
\begin{split}
\abs{\Tr(C^\dagger U)}^2 &= d^2 \Tr(J(\calC) J(\calU)) \\
&= d^2 \bigl\lvert \bra{\Phi^+} (C^\dagger \tensr I) (U \tensr I) \ket{\Phi^+} \bigr\rvert^2.
\end{split}
\end{equation}

Thus, the quantity $\abs{\Tr(C^\dagger U)}^2$ can be estimated via the following procedure:  (1) prepare a maximally entangled state $\ket{\Phi^+}$ on two registers; (2) apply the unknown unitary operation $U$ (on the first register); (3) apply the prescribed unitary operation $C^\dagger$ (on the first register); (4) measure both registers in the Bell bases; (5) repeat the above steps, and count the number of times that the outcome $\ket{\Phi^+}$ is observed.

The second way of estimating $\abs{\Tr(C^\dagger U)}^2$ makes use of a more sophisticated technique known as randomized benchmarking tomography \cite{K14, JSR15}.  This method is robust to SPAM errors, but it is also quite resource-intensive.  In addition, this method requires that the measurement matrix $C$ be chosen from some group that has a computationally efficient description, such as the Clifford group \cite{Gott97, DLT02}.

In randomized benchmarking tomography, one implements a sequence that alternates between applying a random Clifford and applying the unknown unitary map $\calU$. This alternation repeats $L$ times, and then is followed by a single recovery Clifford operation. For example, the recovery operation could be the Clifford that inverts the action of the $L$ randomly applied Cliffords in the sequence, so the total effect of the $L+1$ applied Cliffords (ignoring the applications of $\calU$) would be the identity. 

This protocol, when repeated many times, and for many different lengths $L$, produces a measurement signature of a decaying exponential in $L$, where the rate of decay depends only on $\Tr(\calU)$. By choosing different recovery operations, one can cause the rate of decay to depend on $\Tr(\calU\calC^\dagger )$ for any Clifford $C$. (This analysis assumes that Clifford operations can be implemented perfectly. If the Clifford operations contain errors, this procedure still works, and it produces a characterization of the process $\calU\circ \Lambda$, where $\Lambda$ is the average Clifford error.)

To summarize, randomized benchmarking tomography allows us to estimate quantities of the form
$\Tr(\calU\calC^\dagger )$, where we can choose $\calC: \rho \mapsto C\rho
C^\dagger$ to be any Clifford operation. We can rewrite
this as
\begin{equation}
\label{eqn-oyster}
\Tr(\calU \calC^\dagger) = d^2 \Tr(J(\calU) J(\calC)) = \abs{\Tr(C^\dagger U)}^2.
\end{equation}

%%%%%%%%%%%%%%%%%%%%%%%%%%%%%%%%%%%%%%%%%%%%%%%%%%%%

\subsection{Numerical Simulation}\label{sec:numerics}

We ran numerical simulations to assess the performance of the PhaseLift algorithm for reconstructing an unknown unitary operation $\calU:\: \rho \mapsto U\rho U^\dagger$ (where $U \in \CC^{d\times d}$), using random Clifford measurements $C_1,\ldots,C_m \in \CC^{d\times d}$.  In our simulations, we chose $U$ at random, by sampling from the Haar distribution on the unitary group.  We then sampled the $C_i$ from the uniform distribution on the Clifford group.  We compared this with a second scenario, where the $C_i$ were sampled from the Haar distribution on the unitary group, since this is the ``best'' measurement ensemble for phase retrieval.

We simulated the measurement procedure (\ref{eqn-silver-meas}) on $\calU$, in the noiseless case ($\eps=0$).  Then we solved the PhaseLift convex program (\ref{eqn-silver}), again in the noiseless case ($\eta=0$), in order to obtain an estimate $\calU'$ of $\calU$.  Here we omitted the non-spikiness constraints (\ref{eqn-silver-nonspiky}), in order to make the convex program easier to solve.  Finally, we computed the reconstruction error in the Frobenius norm, $\norm{J(\calU') - J(\calU)}_F$.

%In this section, we discuss numerical simulation of our procedure for unitary
%reconstruction. While we would like to solve the semidefinite program described in Eq. (\ref{eqn-convex-II}), the non-spiky constraint is difficult to implement. Therefore, in order to make the numerics feasible, we instead look at the following problem. 

%Let $\mathcal G$ be a collection of unitaries (we will consider the two cases of Clifford unitaries and Haar random unitaries). Choose $m$ elements $ D_1,\dots, D_m\in \mathcal G$. Let $\sop D_i$ be the CPTP map corresponding to $D_i$. Let $\sop U$ be the map correspondnig to a Haar random unitary $U$. Then let 
%\begin{align}
%A(\rho)&=\left[\tr(\rho J(\sop D_i)\right]_{i=1}^m\nonumber\\
%y&=\left[\tr(J(\sop U) J(\sop D_i)\right]_{i=1}^m.
%\end{align}
%We solve the following problem
%\begin{equation}
%\label{eqn-convex-numeric}
%\begin{split}
%\arg &\min_{\rho \in \CC^{d^2\times d^2}_{\Herm}} \Tr(\rho) \text{ such that} \\
%&\norm{\tilA(\rho) - y}_2 =0, \\
%&\rho \succeq 0, \\
%&\Tr_1(\rho) = \Tr(\rho)\, \I/d, \\
%&\Tr_2(\rho) = \Tr(\rho)\, \I/d.
%\end{split}
%\end{equation}

\begin{figure}[h!]
\includegraphics[width=.48\textwidth]{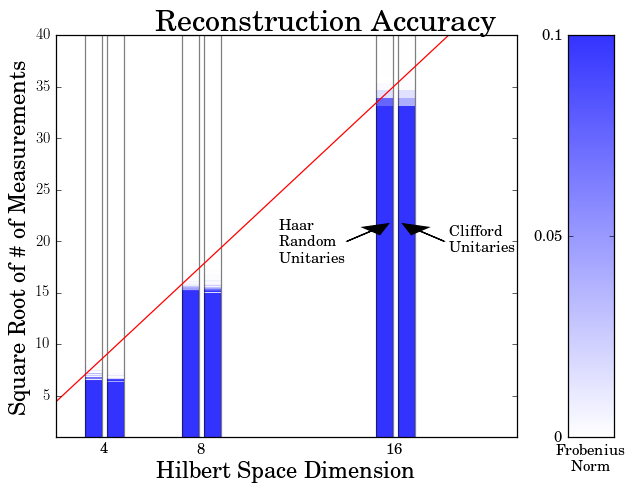}
\caption{\label{fig:simulation} 
Phase retrieval of an unknown unitary matrix $U \in \CC^{d\times d}$, 
using measurement matrices $C_1,\ldots,C_m \in \CC^{d\times d}$ 
which are chosen to be either Haar random unitaries or random Cliffords. 
The $x$-axis shows the dimension $d$, and the $y$-axis shows the square root 
$\sqrt{m}$ of the number of measurements.
For each choice of dimension $d \in \set{2^2, 2^3, 2^4}$, the two vertical bars 
correspond to the two cases where the measurement matrices are either 
Haar random unitaries or random Cliffords. 
The color scale shows the error of the reconstructed matrix $U'$, measured in Frobenius norm; 
white denotes near-perfect reconstruction. 
Each data point was averaged over 10 runs, where each run used fresh random samples of 
$U$ and $C_1,\ldots,C_m$.
The red line shows the expected linear scaling, $\sqrt{m} = \alpha d$, 
where $\alpha = 2.2$ was chosen to fit the data.
Somewhat surprisingly, random Clifford measurements perform nearly as well as 
measurements using Haar-random unitaries. 
%Accuracy of unitary reconstruction
%using either Haar random unitaries or random Clifford unitaries. The $x$-axis
%gives the dimension of the Hilbert space on which the unitaries act. The $y$-axis is the square root of the number of measurements used in the reconstruction.
%Each point gives the Frobenius norm difference between the true unitary
%and the reconstructed unitary, averaged over 10 runs. In each run, a random
%unitary is chosen as the target, and some number (given by the $y$-axis) of random Haar or Clifford unitaries are chosen and used to reconstruct the original unitary via Eq. (\ref{eqn-convex-numeric}). The left column at each dimension corresponds to Haar random measurements, and the right column corresponds to Clifford measurements. The red line is $2.2d$ (to guide the eye).
}
\end{figure}

The results are shown in Figure \ref{fig:simulation}.  These results suggest that phase retrieval using random Clifford measurements performs nearly as well as phase retrieval using Haar-random unitary measurements.  In both cases, the number of measurements scales as $m \approx C d^2$, where $C \approx 4.8$.  This suggests that, although random Clifford operations are not a unitary 4-design, they are ``close enough'' to ensure accurate reconstruction of unitary matrices via phase retrieval.

%We consider the case that $\mathcal G$ is the set of Clifford unitaries (which
%form a unitary-3 design), and also the case that $\mathcal G$ is the set of
%Haar random unitaries. In Figure \ref{fig:simulation}, we plot the difference
%in Frobenius norm between the solution of  Eq. (\ref{eqn-convex-numeric}) and
%$J(\sop U)$ as a function of number of measurement settings and dimension. We consider unitaries acting on $4,$ $8,$ and $16$ dimensional
%Hilbert spaces. For each dimension, we show the data for Haar random
%measurements in the left column, and Clifford measurements in the right column.
%The $y$-axis gives the square root of the number of measurement settings. 
%Each point is an average of 10 runs, where in each run, a new random unknown
%unitary $\sop U$ is chosen, and $m$ new measurement unitaries are chosen.

%We only ran simulations for even powers because Cliffords are defined on Hilbert spaces whose dimensions are powers of 2. Due to resource constraints, we were not able to go to dimensions larger than $16.$

%To guide the eye, the red line in Figure \ref{fig:simulation} shows the function $2.2 d,$ which suggests the number measurement settings required for reconstruction scales roughly as $d^2$, as expected.

%Surprisingly, the reconstruction using Cliffords does as well in practice as true Haar random unitaries. Thus we expect that the analysis presented in this paper can be improved. 

%%%%%%%%%%%%%%%%%%%%%%%%%%%%%%%%%%%%%%%%%%%%%%%%%%%%%%%%%

\subsection*{Acknowledgments}

The authors thank Richard Kueng and David Gross for helpful discussions about
this problem; Easwar Magesan for sharing his Clifford sampling code; 
Aram Harrow and Steve Flammia for clarifications regarding the results in \cite{Mendl08}; 
and several anonymous reviewers for suggestions which improved this paper. 
Contributions to this work by NIST, an agency of the US
government, are not subject to US copyright.

%%%%%%%%%%%%%%%%%%%%%%%%%%%%%%%%%%%%%%%%%%%%%%%%%%%%%%%%%

%\bibliographystyle{plain}
%\bibliography{cs2designsbib}
%\input{cs2designs2_arxiv_sk0.bbl}

%%%%%%%%%%%%%%%%%%%%%%%%%%%%%%%%%%%%%%%%%%%%%%%%%%%%%%%%%

\appendix

\section{Representations of Quantum Operations}
\label{app:small_calcs}
Given a completely positive and trace preserving quantum operation $\sop F:\mathbb C^{d\times d}
\rightarrow C^{d\times d}$, there are several
useful representation of $\sop F$. 
One is the Choi-Jamiolkowski representation $J(\sop F) \in \CC^{d^2\times d^2}$,
\begin{equation}
J(\sop F) = (\sop F \tensr \calI) (\ket{\Phi^+}\bra{\Phi^+}), 
\end{equation}
which is obtained by applying $\sop F$ to one half of the maximally entangled state 
%\begin{equation}
$\ket{\Phi^+} = \frac{1}{\sqrt{d}} \sum_{i=0}^{d-1} \ket{i} \tensr \ket{i} \in \CC^{d^2}$.
%\end{equation}
Another is the Liouville representation $\sop F^L\in\mathbb{C}^{d^2\times d^2}$, 
\begin{equation}
\begin{split}
\bra{kl}\sop F^L\ket{ij} &= \bra{k}\sop F(\ketbra{i}{j})\ket{l} \\
 &= d \bra{ki} J(\sop F) \ket{lj},
\end{split}
\end{equation}
where $\ket{i},$ $\ket{j}$, $\ket{l}$ and $\ket{k}$
are any standard basis states.

All representations are completely equivalent, and it is
a simple exercise to convert between them. However,
certain representations make it easier to check
for properties like complete positivity. 

For example from the complete positivity constraint, we have
that $J(\sop X)$ is Hermitian, so
\begin{align}
\bra{ki}J(\sop X)\ket{lj}=\bra{lj}J(\sop X)^*\ket{ki}.
\end{align}
Converting this into Liouville representation, we have
\begin{align}
\bra{kl}\sop X^L\ket{ij}=\bra{lk}(\sop X^L)^*\ket{ji}.\label{eq:conj_relation}
\end{align}

Using this fact, we can show that for completely positive
superoperators $\sop F$ and $\sop K$,
\begin{align}\label{eq:conv-reps}
\tr(J(\sop F)J(\sop K))=
\frac{1}{d^2}\tr(\sop F^L(\sop K^L)^\dagger).
\end{align}
To see this, we calculate
\begin{align}
\tr(J(\sop F)J(\sop K))=&\frac{1}{d^2}
\tr\left(\sum_{ij}\sop F(\ketbra{i}{j})
\sop K(\ketbra{j}{i})\right)\nonumber\\
=&\frac{1}{d^2}\sum_{ijkl}\bra{k}\sop F(\ketbra{i}{j})\ketbra{l}{l}\sop K(\ketbra{j}{i})\ket{k}\nonumber\\
=&\frac{1}{d^2}\sum_{ijkl}\bra{kl}\sop F^L\ket{ij}\bra{kl}(\sop K^L)^*\ket{ij}\nonumber\\
=&\frac{1}{d^2}\tr(\sop F^L(\sop K^L)^\dagger).
\end{align}

Additionally, we note that
for unitary maps $\calU:\: \rho \mapsto U\rho U^\dagger$, where $U$ is 
a unitary operation, the corresponding 
Liouville representation takes the form 
\begin{align}\label{eq:lrep}
\sop U^L=U\otimes U^*,
\end{align}
because
\begin{align}
\sop U(\ketbra{i}{j})=U\ketbra{i}{j}U^\dagger
\end{align}

Using Eq. (\ref{eq:conv-reps}) and Eq. (\ref{eq:lrep}), we have that
for a map $\sop U$ representing a unitary $U$ and a map $\sop
C$ representing a unitary $C,$   that 
\begin{align}
\Tr(J(\calU) J(\calC))
= (1/d^2) \abs{\Tr(U^\dagger C)}^2.
\end{align}

%%%%%%%%%%%%%%%%%%%%%%%%%%%%%%%%%%%%%%%%%%%%%%%%%

\end{document}